\documentclass[prd,twocolumn,superscriptaddress,altaffilletter,showpacs]{revtex4}
\usepackage{amssymb,amsmath}
\usepackage{graphicx}

\usepackage{tikz}
\usetikzlibrary{arrows,decorations.pathmorphing}
\usetikzlibrary{decorations.text}

\def\be{\begin{equation}}
\def\ee{\end{equation}}
\def\bea{\begin{eqnarray}}
\def\eea{\end{eqnarray}}

\usepackage{xcolor}
\definecolor{ao(english)}{rgb}{0.0, 0.5, 0.0}

\begin{document}

\title{Mass and spin of a Kerr black hole in modified gravity and a test of the Kerr black hole hypothesis}

\author{Pankaj Sheoran} \email{hukmipankaj@gmail.com}
\affiliation{National Institute of Technology (N.I.T), Calicut, Kerala-673601, India.}

\author{Alfredo Herrera-Aguilar} \email{aherrera@ifuap.buap.mx}
\affiliation{Instituto de F\'{\i}sica, Benem\'erita Universidad Aut\'onoma de Puebla,\\ Apartado Postal J-48, C.P. 72570, Puebla, Puebla, M\'{e}xico.}
\affiliation{Instituto de F\'{\i}sica y Matem\'{a}ticas, Universidad Michoacana de San Nicol\'{a}s de Hidalgo.\\ Edificio C-3, 58040 Morelia,
Michoac\'{a}n, M\'{e}xico}

\author{Ulises Nucamendi} \email{ulises@ifm.umich.mx}
\affiliation{Instituto de F\'{\i}sica y Matem\'{a}ticas, Universidad Michoacana de San Nicol\'{a}s de Hidalgo,\\ Edificio C-3, 58040 Morelia,
Michoac\'{a}n, M\'{e}xico}

\date{\today}

\begin{abstract}

In this paper we compute the  Arnowitt-Deser-Misner (ADM) mass, the angular momentum and the charge of the Kerr black hole solution in the scalar-tensor-vector gravity theory [known as the Kerr-MOG (modified-gravity) black hole configuration]; we study in detail as well several properties of this solution such as the stationary limit surface, the event horizon, and the ergosphere, and conclude that the new deformation parameter $\alpha$ affects the geometry of the Kerr-MOG black hole significantly in addition to the ADM mass and spin parameters. Moreover, the ADM mass and black hole event horizon definitions allow us to set  a novel upper bound on the deformation parameter and to reveal the correct upper bound on the black hole spin. We further find the geodesics of motion of stars and photons around the Kerr-MOG black hole. By using them we reveal the expressions for the mass and the rotation parameter of the Kerr-MOG black hole in terms of the red- and blueshifts of photons emitted by geodesic particles, i.e., by stars. These calculations supply a new and simple method to further test the general theory of relativity in its strong field limit: If the measured red- and blueshifts of photons exceed the bounds imposed by the general theory of relativity, then the black hole is not of Kerr type. It could also happen that the measurements are allowed by the Kerr-MOG metric, implying that the correct description of the dynamics of stars around a given black hole should be performed using MOG or another modified theory of gravity that correctly predicts the observations. In particular, this method can be applied to test the nature of the putative black hole hosted at the center of the Milky Way in the near future.

\end{abstract}


\maketitle


\section{Introduction}

The presence of a great number of extremely compact and massive objects in our Universe has led to the hypothesis that they are black holes (BHs). Current and near-future precise observations of the orbital motion near these black hole candidates have the potential to determine if they possess the spacetime structure predicted by the general theory of relativity (GTR), providing a test of the theory in its strong gravitational regime.

Recently, a method to compute the spin $a$ and the mass $M$ parameters of the Kerr black hole, which is an exact solution of Einstein's field equations of the GTR, in terms of red- and blueshifts ($z_{r}$ and $z_{b}$) of photons emitted by massive geodesic objects was proposed in \cite{HN_2015}. The authors have shown that in principle with an observational data set of these red- and blueshifts of photons emitted by stars orbiting in the background of the Kerr spacetime with different radii, one can easily compute the mass $M$ and the spin $a=J/M$ parameters of the Kerr BH (here $J$ stands for the BH angular momentum).

This method has been applied as well to a higher-dimensional Myers-Perry BH in \cite{Sharif-Iftikhar} and to the Kerr-Sen BH arising in the heterotic string theory in \cite{Kuniyaletal}; it was further used to derive mass formulas for a Reissner-Nordstrom BH and boson stars in \cite{Ulises} and for a noncommutative geometry inspired a Schwarzschild BH in \cite{Uniyaletal}. A similar approach was implemented to model a binary system in the external gravitational field of a Schwarzschild BH in order to fit the timing data from the x-ray pulsars that move in the neighborhood of our Galactic Center (Sgr A*) in \cite{GKT} (for a review regarding the observations of Sgt A* and processes around it in the electromagnetic spectrum see \cite{Goddietal}).

It turns out that most of the tests of the GTR, which is considered as one of the most beautiful of all existing physical theories, have been performed in the weak field regime \cite{Will_1993, Will_2014, Stairs_2003, Wex_2014, Murata_2015}; however, the recent detection of gravitational waves produced by coalescing black holes has opened a new era in the study of gravity in the strong coupling regime  \cite{gw1}-\cite{gw5}. Therefore, tests of the GTR are important in the strong gravitational limit; these tests will also probe modified theories of gravity that can show a significant departure from GTR only in such a regime. And, to probe the strong gravity regime, BHs are the most promising candidates because of their large gravitational field near the event horizon.

There have been several generalizations of rotating BH solutions both within the GTR and within modified gravitational theories. Among them we can mention the generalization of the Kerr-Newman spacetime possessing a full set of mass-multipole momenta which describes the exterior gravitational field of a charged rotating arbitrary axisymmetric mass \cite{MN} (see as well \cite{VH} for the bumpy Kerr BH version). A study of prolate/oblate deformations of Kerr spacetimes using this family of metrics was performed in \cite{GLM}. The so-called quasi-Kerr metric \cite{GB} that incorporates just one independent quadrupole moment has been used to propose a test of the no-hair theorem in \cite{JP1}. Thus, these deformed BHs differ from the Kerr metric in (at least) one multipole moment and allow for a test that distinguishes a Kerr BH from an object of a different kind. Moreover, a deviation of the quadrupole moment from the predicted Kerr BH value will lead to either prolate or oblate images of BHs, depending on the sign and magnitude of the measured deviation. In particular, there will appear changes in the structure of the BH shadow \cite{JP2}.

On the other hand, BHs are supposed to be hosted at the center of every galaxy, including our own Milky Way. Therefore, one can compute observable quantities that can be measured directly with either current or near-future instruments. These observational experiments allow for the formulation of a GTR test in the strong gravity regime: If the central object of a galaxy is a black hole, there must be no deviation from the Kerr metric. If, however, the deviation is measured to be different from zero, then it indicates that either the central object is of a different type, or that the GTR itself breaks down at the strong gravitational regime very close to the event horizon of the BH. However, one should always keep in mind that the Kerr BH solution to the GTR field equations is indistinguishable from exact solutions of a wide variety of gravity theories that add dynamical vector and tensor degrees of freedom to the Einstein-Hilbert action \cite{Psaltisetal}.

An interesting family of metrics that contain an infinite amount of parameters, is regular everywhere outside of the event horizon, and predicts a different shadow from that expected from the GTR Kerr BH solution was proposed in \cite{JP3}; a simplified version of this general metric that contains just three parameters can be found in \cite{JohannsenThreeConstanst2013}. In these metrics, circular equatorial orbits of massive and massless test particles around the BH, as well as innermost stable circular orbits, significantly change for even moderate deviations from the Kerr BH metric, providing a suitable arena for carrying out strong gravitational field tests of the GTR.

The above-mentioned phenomenological modifications of the Kerr BH metric have been supplemented by several exact solutions to the field equations of modified gravitational theories. For instance, a generalization of the bumpy Kerr BH solution \cite{VH} has been mapped to known analytical BH solutions in alternative theories of gravity in \cite{VYS}; within the context of the $f(R)$ gravity theory rotating BH configurations have been constructed in \cite{larranaga}-\cite{cembranos}; in dilatonic and axidilatonic string theories, rotating BH solutions were obtained in \cite{Sen} and \cite{GGK}, respectively; in the Einstein-Dilaton-Gauss-Bonnet (EDGB) theory with large coupling constants a rotating  BH was reported in \cite{CPR}; within the scalar-tensor-vector modified gravitational (MOG) theory, a new class of rotating BHs (known as Kerr-MOG BHs) was proposed in \cite{Moffat_2015}. The latter Kerr-MOG BHs are defined by the spin parameter $a$, the mass parameter $M_{\alpha}$, and a deformation parameter $\alpha$
\cite{footnote1},
in contrast to the Kerr BH which is only defined by the spin $a$ and the mass $M$ parameters.
This family of metrics produces a photon sphere which observationally differs from the one generated by the pure Kerr BH \cite{Moffat2}. Moreover, gravitational lensing of the BH and the images of the BH shadow provide observational signatures for distinguishing between Kerr and Kerr-MOG BH configurations.

An interesting parametrization for a general stationary and axisymmetric BH metric has been proposed in \cite{KRZh}. With a small number of parameters, this formalism yields several known rotating BH configurations, among them we find the Kerr BH, the rotating dilaton (or Kerr-Sen) and the EDGB BHs, and the Johannsen-Psaltis metric \cite{JP3}. On the basis of this parametrization, the shadow analysis of various BHs in alternative theories of gravity was performed in \cite{YZhRKM}.

Most of the aforementioned BH metrics allow, in principle, for a test that distinguishes a Kerr BH from a different compact object. For interesting reviews on the subject see \cite{Psaltis}-\cite{Bambi2}.

Many other tests of GTR have been proposed. A study of the evolution of the spin parameter of accreting compact objects with non-Kerr quadrupole moment showed that if these supermassive objects are not Kerr BHs, the accretion process can make them reach a superspinning regime \cite{Bambi1}. The profiles of fluorescent broadened iron lines emitted from the accretion flows around a BH were calculated as a function of its mass, spin parameter, and a free parameter that measures potential deviations from the Kerr metric in \cite{JP4}. For complete reviews of different tests of the GTR in the strong gravitational field regime we refer to \cite{Bertietal}- \cite{YS}.

In this paper we obtain the following new results: We compute the ADM mass for the Kerr-MOG metric. We establish upper bounds on both  the deformation and spin parameters of the Kerr-MOG spacetime (the ADM mass definition leads to a novel upper bound on the deformation parameter and reveals the correct upper bound on the black hole spin parameter). We reveal the correct structure of the ergosphere of the Kerr-MOG BH on the basis of the ADM mass definition since it changes the expressions of both the stationary limit surface and the event horizon, yielding interesting effects that differ from those predicted by Moffat's original metric. We obtain a correct definition of the extremal limit for the Kerr-MOG BH metric, allowing us to correctly compute the red- and blueshifts of photons in subsequent sections.
We generalize the method for computing the spin and mass parameters of the Kerr-MOG BH in terms of red- and blueshifts of photons emitted by massive geodesic particles, in the spirit of \cite{HN_2015}. We also provide a formulation of a novel concrete test of the Kerr BH hypothesis on the basis of these red- and blueshifts of photons: given an observational data set of red- and blueshifts of photons emitted by stars orbiting around a rotating BH with different radii, one can, in principle, determine whether the spin and mass parameters correspond to the predicted values of the Kerr BH of the GTR, to the Kerr-MOG BH of the scalar-tensor-vector gravity(STVG) modified gravitational theory for a given value of the $\alpha$ parameter, or correspond to a BH or compact object of a different theory, providing a further test of the GTR in the strong gravitational regime.
We finally provide new algebraic expressions for estimating the mass and spin parameters of the Kerr-MOG BH from observational data; these
expressions radically differ from those found for the Kerr metric, in particular, the order of the algebraic equations for $a$ and $M_{\alpha}$ is doubled for the Kerr-MOG BH in comparison to the Kerr BH.

Thus, in Sec. II we shall start with the line element of the Kerr-MOG BH and we shall compute its ADM mass, its angular momentum, and its charge; we shall also study the behavior of its stationary limit surface (SLS), its event horizon (EH), and its ergosphere with emphasis on the variation along the $\alpha$ parameter; with the expressions for the ADM and event horizon at hand we set an upper bound on the deformation parameter and show that the spin parameter is bounded from above as for the Kerr BH of the GTR.
In Sec. III, we shall compute the geodesics of both massive and massless particles in the background of a Kerr-MOG BH and particularize them for the equatorial (i.e., in the $\theta=\pi/2$ plane) and circular cases. Further, in Sec. IV we shall discuss the red- and blueshifts of photons emitted by the geodesic particles orbiting around the Kerr-MOG BH, whereas in Sec. V, we set bounds on these red- and blueshifts in terms of the spin parameter and formulate a novel possible test for the Kerr BH hypothesis; in Sec. VI we shall find the expressions for the spin $a$ and mass $M_{\alpha}$ parameters of the Kerr-MOG BH in terms of these red- and blueshifts, and finally, we shall end the paper by summarizing our results and evoking some future prospects in Sec. VII.

\vspace{-0.2cm}

\section{The Kerr-MOG black hole metric}

The field equations for scalar-tensor-vector gravity (also known as MOG in the literature) have a stationary, axisymmetric black hole solution named the Kerr-MOG black hole, which is determined by its mass, its angular momentum, and a deformation parameter $\alpha$ \cite{Moffat_2015}.
The Kerr-MOG BH metric of modified gravity in Boyer-Lindquist coordinates with $c=1$ reads
\begin{eqnarray}
\label{metric}
       ds^2 &=& -\left(\frac{-a^2\sin^{2}\theta+\Delta}{\Sigma}\right) dt^2
        +\frac{\Sigma}{\Delta}dr^{2}\nonumber\\
        &-&2\left(\frac{r^2+a^2-\Delta}{\Sigma}\right)a \sin^{2}\theta dt d\phi
      +\Sigma d\theta^2\nonumber\\
       &+&\left[\frac{-\Delta a^{2}\sin^2\theta +\left(r^2+a^2\right)^2}{\Sigma}\right]\sin^2\theta d\phi^2,
\end{eqnarray}
where
\begin{eqnarray}
 \Delta&=&r^2-2GMr+a^2+M^2\alpha\, G\, G_N,\nonumber\\
 \Sigma&=&r^2+a^2\cos^2\theta,
\end{eqnarray}
where $G=G_{N}(1+\alpha)$ is an enhanced gravitational constant defined with the aid of Newton's gravitational constant $G_{N}$ and the deformation parameter $\alpha=\frac{(G-G_N)}{G_N}$ introduced in \cite{Moffat_2013}.

At this point it is worth remembering that a proportionality relation between the charge $Q$ of the MOG vector field and the mass parameter $M$ of the metric (\ref{metric}) was postulated  in \cite{Moffat_2015}:
\begin{equation}
\label{QM}
Q=\sqrt{\alpha G_N}M,
\end{equation}
endowing the charge $Q$ with a gravitational character, since astrophysical bodies, including black holes are electrically neutral. This gravitational charge yields the modified Newtonian acceleration that is used to fit galaxy rotation curves, galaxy cluster dynamics as well as solar system and binary pulsar dynamics \cite{Moffat2006}. From now onwards, throughout this paper, we are taking $G_{N}=1$.

The metric (\ref{metric}) is stationary and axially symmetric. We denote by $\xi$ and $\psi$ the Killing vector fields which are the generators of the corresponding symmetry transformations
\begin{eqnarray}
\label{KillingT}
  \xi^{i} &=& (1,0,0,0)   \quad \textrm{timelike Killing vector field,}  \\
  \label{KillingR}
  \psi^{i} &=& (0,0,0,1)  \quad \textrm{rotational Killing vector field.}
\end{eqnarray}

The presence of these Killing vectors is very useful when computing the invariant parameters that define the Kerr-MOG metric (\ref{metric}): the ADM mass, the angular momentum and the total charge \cite{Wald}, namely
\begin{equation}
\label{ADMmass}
M_{ADM}=-\frac{1}{8\pi}\int_S\epsilon_{ijkl}\nabla^{k}\xi^{l}=(1+\alpha)M\equiv M_{\alpha}\,,
\end{equation}
\begin{equation}
\label{J}
J=\frac{1}{16\pi}\int_S\epsilon_{ijkl}\nabla^{k}\psi^{l}=M(1+\alpha)\,a\equiv M_{\alpha}\,a\,,
\end{equation}
\begin{equation}
\label{charge}
4\pi Q=\frac{1}{2}\int_S\epsilon_{ijkl}B^{kl},
\end{equation}
where $S$ is an asymptotic 2-sphere and $B_{kl}=\partial_k\phi_l-\partial_l\phi_k$ is the strength tensor of the vector field $\phi_k$ of the MOG theory \cite{Moffat_2013}.

Thus, we should stress here that the ADM mass $M_{\alpha}$ is the gravitational mass of the Kerr-MOG BH since it corresponds to the value of the energy when considering the Hamiltonian formalism of the GTR
\cite{footnote2}.
Therefore, we shall further denote the Kerr-MOG BH mass as $M_{\alpha}$ since this quantity is precisely the source of the gravitational force acting on a test particle (i.e., a star, for instance) moving around the black hole.

On the other hand, we know that, unlike the Schwarzschild BH, the Kerr BH has two important surfaces: the SLS and the event horizon one. Hence, it is interesting to calculate and study these surfaces for the metric (\ref{metric}). At the SLS, the Killing vector $\xi$ becomes null which further implies that the prefactor of $dt^2$ (which determines the rate of flow of time) vanishes. The equation of this surface is
\begin{equation}
\label{sls}
 a^2\sin^{2}\theta-\Delta=0.
\end{equation}
It can be also written in explicit form $r=r_{SLS}$, where $r_{SLS}$ is given by the relation
\begin{eqnarray}
\label{sls1}
r_{SLS}= M_{\alpha} \pm \sqrt{\frac{M_{\alpha}^2}{(1+ \alpha)}-a^2\cos^{2}\theta}.
\end{eqnarray}

\begin{figure*}
\begin{tabular}{c c}
\includegraphics[scale=0.55]{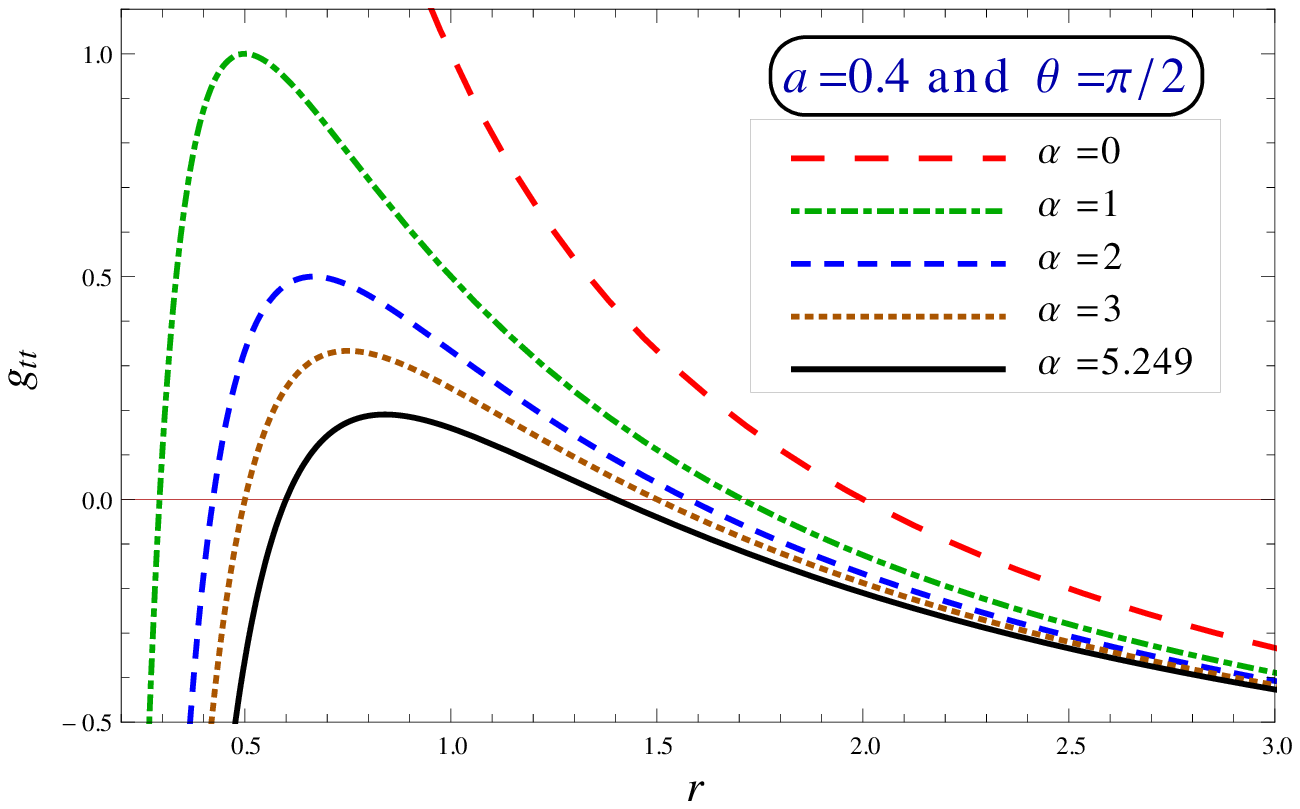}\hspace{0cm}
&\includegraphics[scale=0.55]{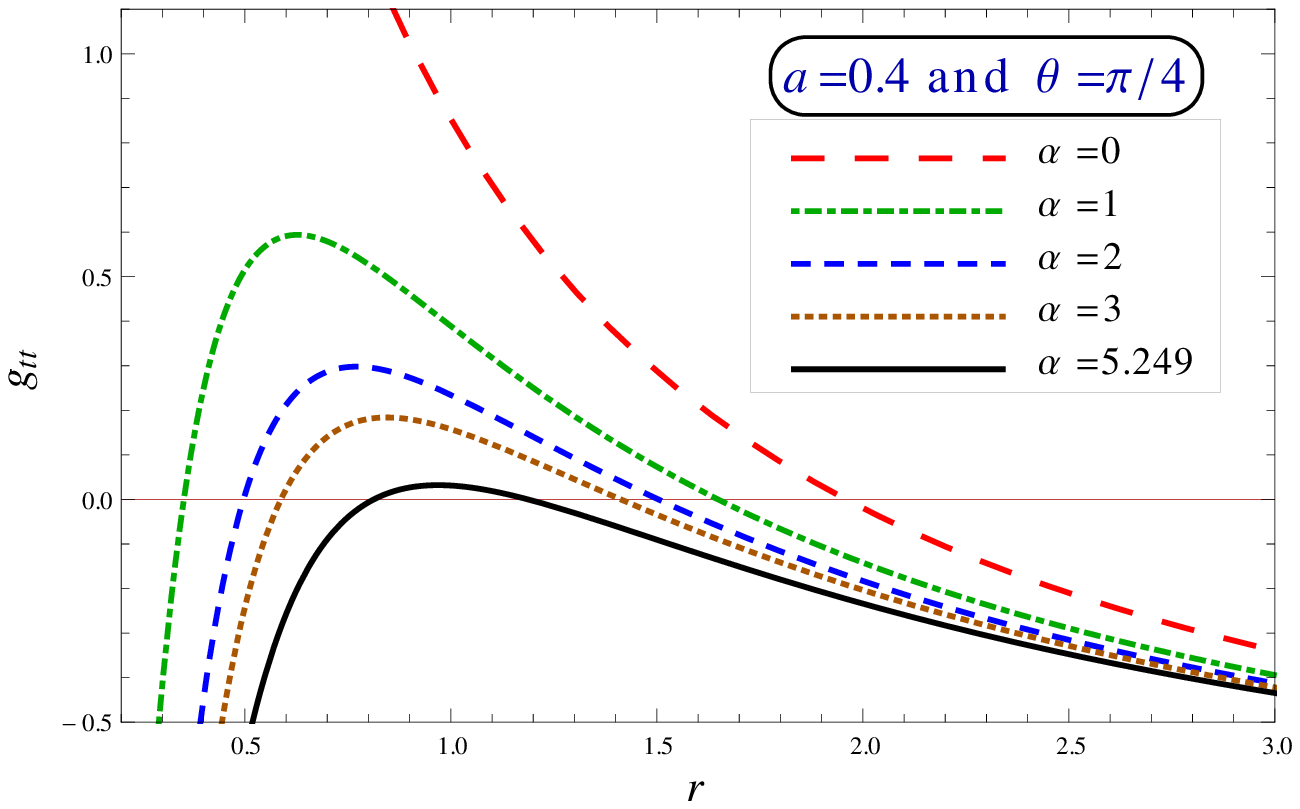}
 \\
 \includegraphics[scale=0.55]{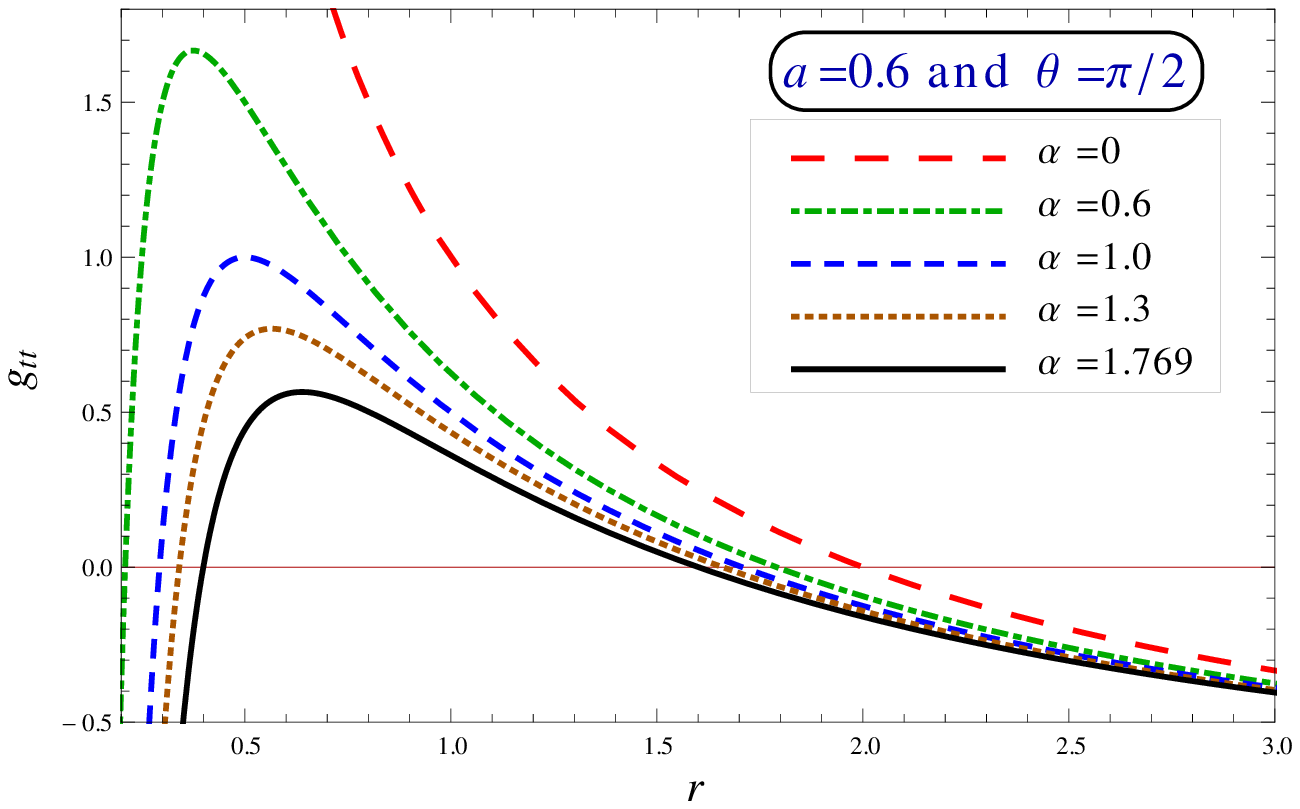}\hspace{0cm}
 &\includegraphics[scale=0.55]{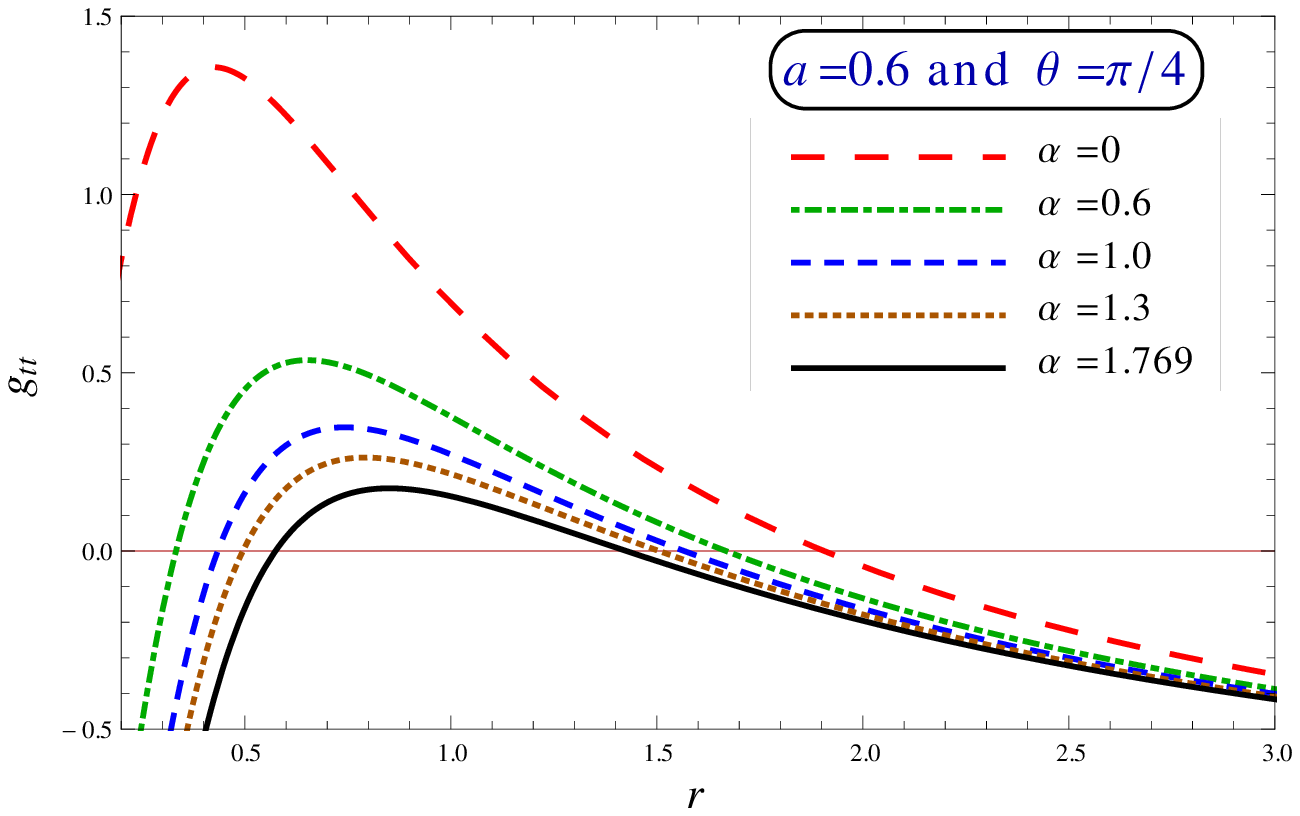}
 \\
 \includegraphics[scale=0.55]{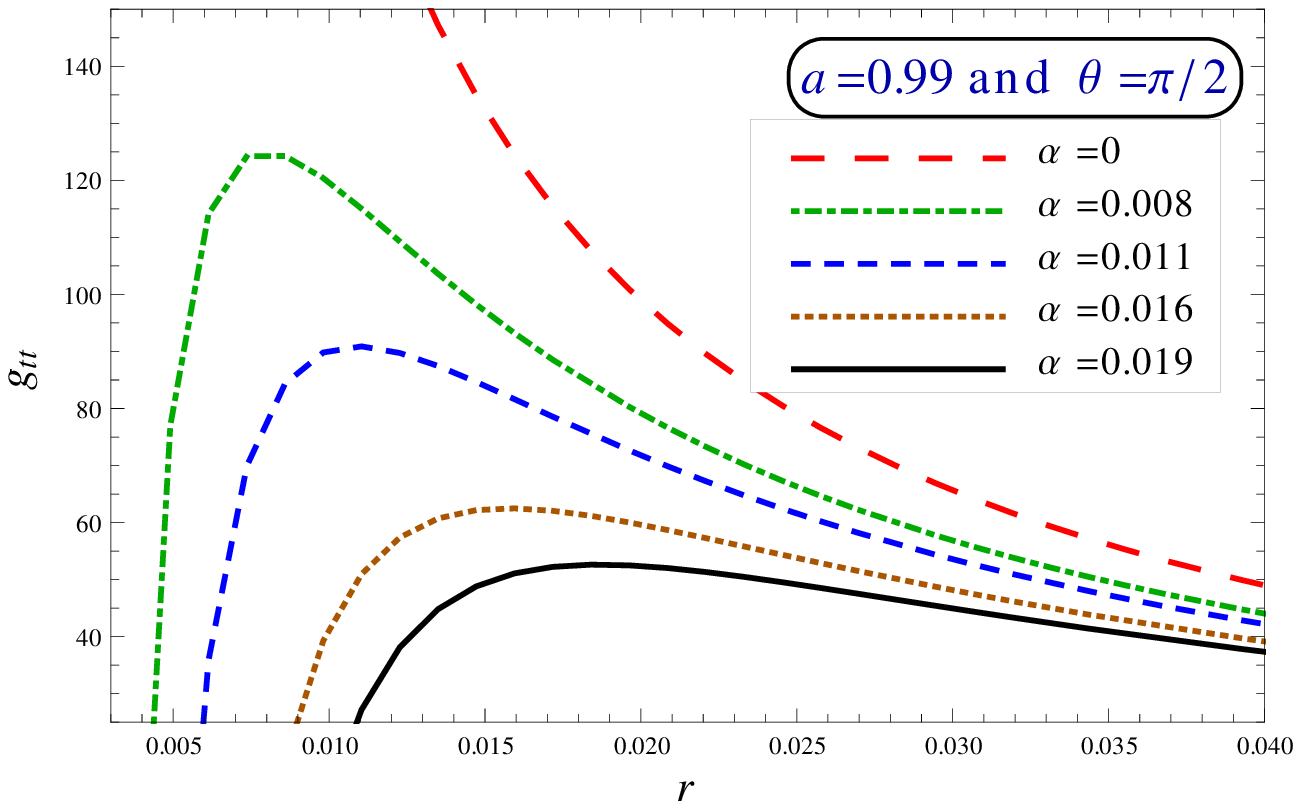}\hspace{0cm}
 &\includegraphics[scale=0.55]{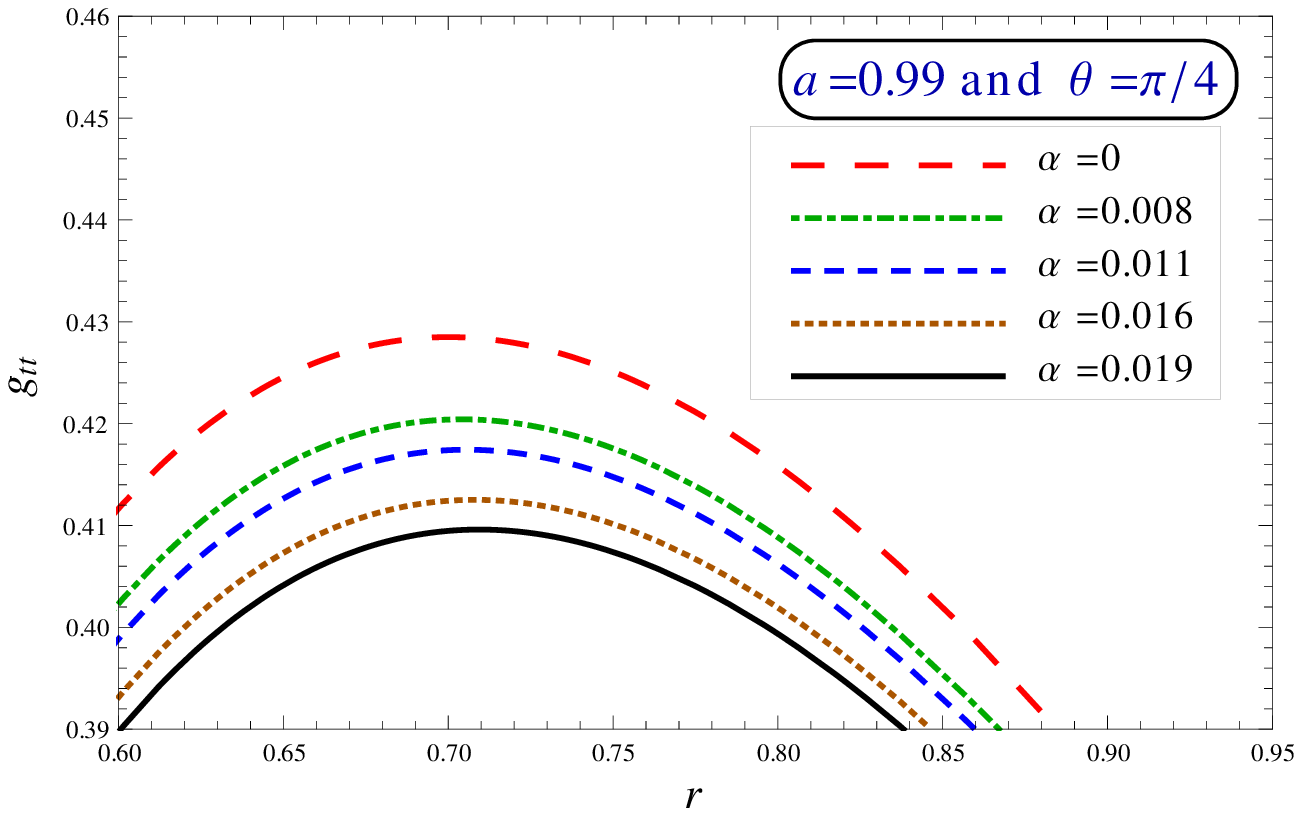}
 \end{tabular}
 \caption{The behavior of the stationary limit surface ($g_{tt}=0$) vs $r$ for different values of $\alpha$ is shown at different $\theta$'s (for $M_{\alpha}=1$). Here, the solid (black) line corresponds to the extremal Kerr-MOG black hole and the large-dashed (red) line to the Kerr BH (i.e., when $\alpha=0$).}
 \label{fig1_gtt}
\end{figure*}

The horizons of the BH (\ref{metric}) are calculated by equating the $g^{rr}$ to zero, i.e.
\begin{equation}
 \label{Delta}
 \Delta=r^2-2M_{\alpha} r + a^2 + \frac{\alpha}{(1+\alpha)}M_{\alpha}^2=0.
\end{equation}
This equation defines two event horizons which lie at $r=r_{\pm}=r_{EH}$, where
\begin{equation}
 \label{EH1}
 r_{EH}=M_{\alpha} \pm \sqrt{\frac{M_{\alpha}^2}{(1+ \alpha)}-a^2}.
\end{equation}

Here it is worth noticing that both the positiveness of the radicand in (\ref{QM}) and the positiveness of the radicand in Eq. (\ref{EH1}) impose physical bounds on the $\alpha$ parameter
\begin{equation}
 \label{alphabounds1}
0\le\alpha\le \frac{M_{\alpha}^2}{a^2}-1,
\end{equation}
inequalities that correspond to a black hole configuration. If the second inequality is inverted, we obtain a naked singularity; however, if it holds,
then $|a|<M_{\alpha}$, a relation which is valid as well for the Kerr BH of the GTR. In fact, the following upper bound can be established for the spin parameter:
\begin{equation}
 \label{abound}
|a|\le \frac{M_{\alpha}}{\sqrt{1+ \alpha}}
\end{equation}
a relation that tells us that the spin parameter of the Kerr-MOG BH will be more restricted with respect to its ADM mass than the rotation parameter of the Kerr BH since the deformation parameter $\alpha$ is positive.

Thus, the correct definition of the ADM mass for the Kerr-MOG BH influences the correct determination of the SLS, the EH, and the ergosphere.

\begin{table}
\begin{center}
\caption{The range of the deformation parameter $\alpha$ corresponding to different values of the spin parameter $a$ is shown for the Kerr-MOG BH. Here the value of the mass parameter $M_{\alpha}$ is unity.}\label{table1}
\vspace{0.5cm}
\begin{tabular}{l c l c l}
 \hline \hline
No. & & $a$  \;\;  & & Range of $\alpha$ \\
\hline

1    &  & 0.3   \;\;  & & $0\leq\alpha\leq 10.111$  \\
2    &  & 0.4  \;\;   & & $0\leq\alpha\leq 5.250$  \\
3    &  & 0.5  \;\;   & & $0\leq\alpha\leq 3.0$  \\
4   &  & 0.6  \;\;   & & $0\leq\alpha\leq 1.777$  \\
5   &  & 0.7  \;\;   & & $0\leq\alpha\leq 1.040$  \\
6   &  & 0.8  \;\;   & & $0\leq\alpha\leq 0.562$  \\
7   &  & 0.9  \;\;   & & $0\leq\alpha\leq 0.234$  \\
8   &  & 0.99  \;\;   & & $0\leq\alpha\leq 0.020$  \\
\hline \hline
\end{tabular}
\end{center}
\end{table}

\begin{figure*}
\begin{tabular}{c c}
\includegraphics[scale=0.60]{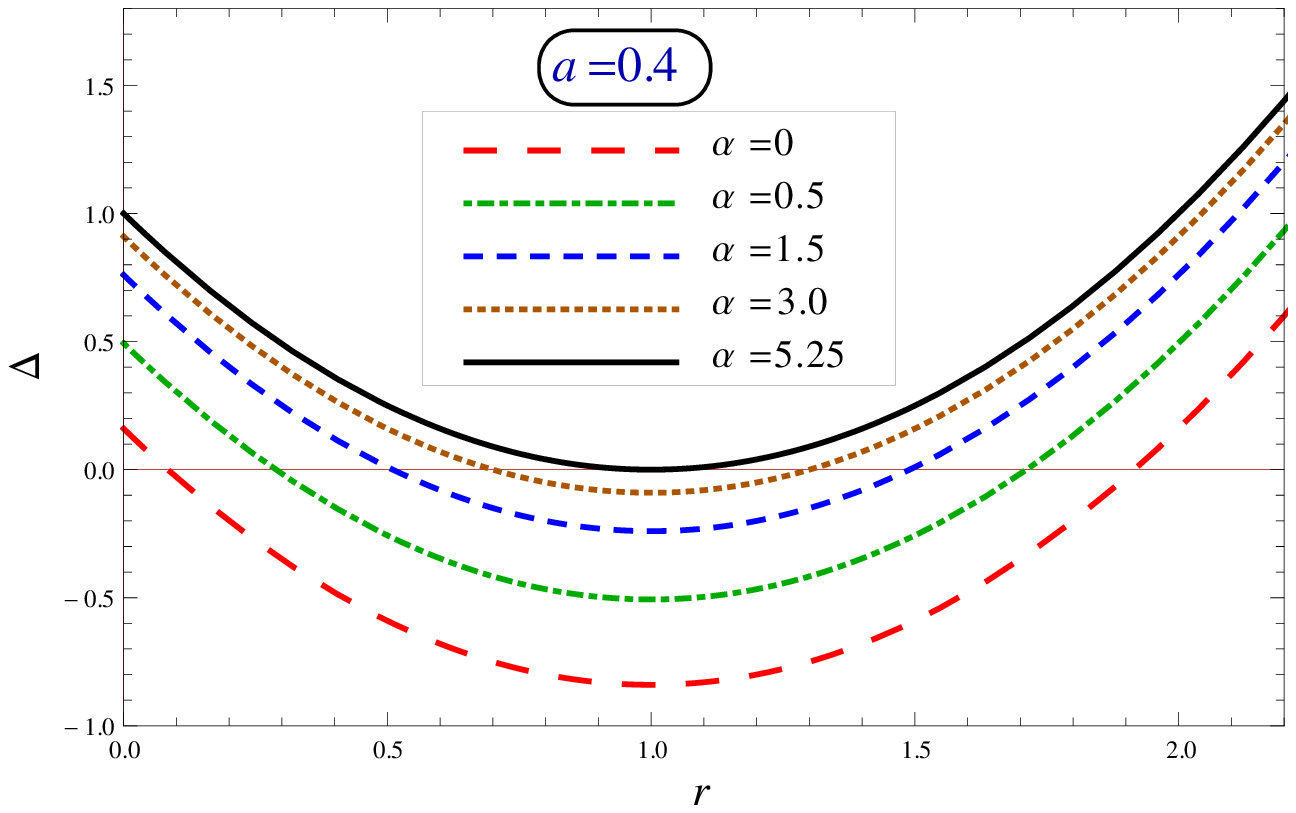}\hspace{0cm}
&\includegraphics[scale=0.60]{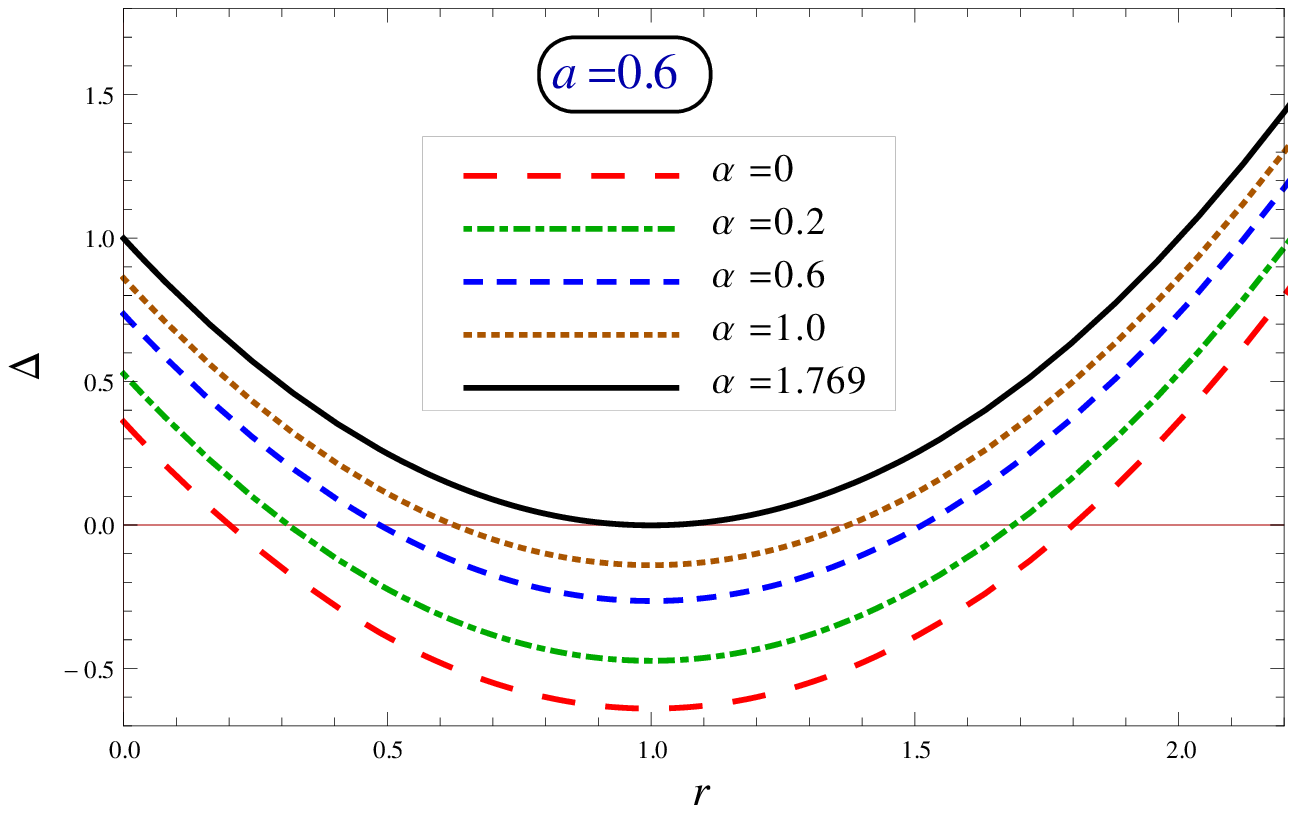}
\\
\includegraphics[scale=0.60]{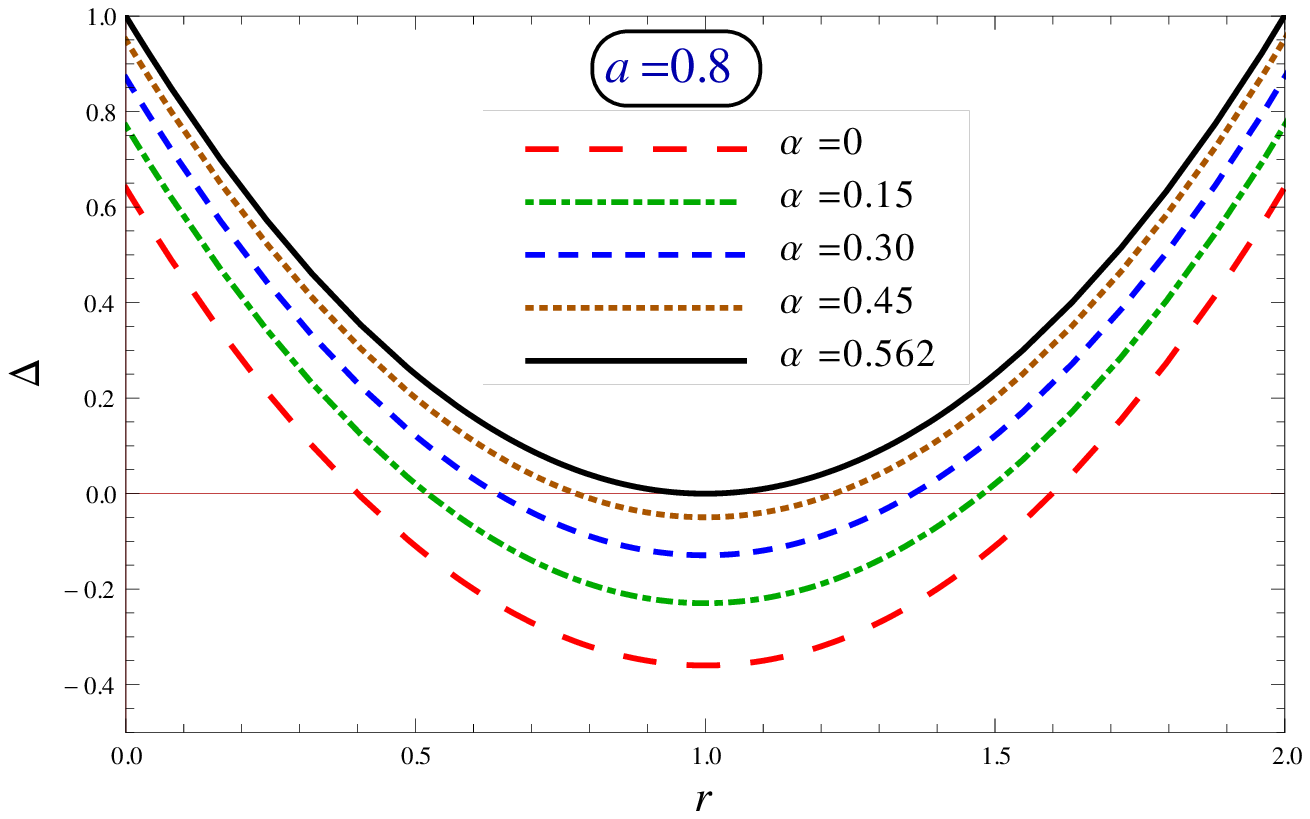}\hspace{0cm}
&\includegraphics[scale=0.60]{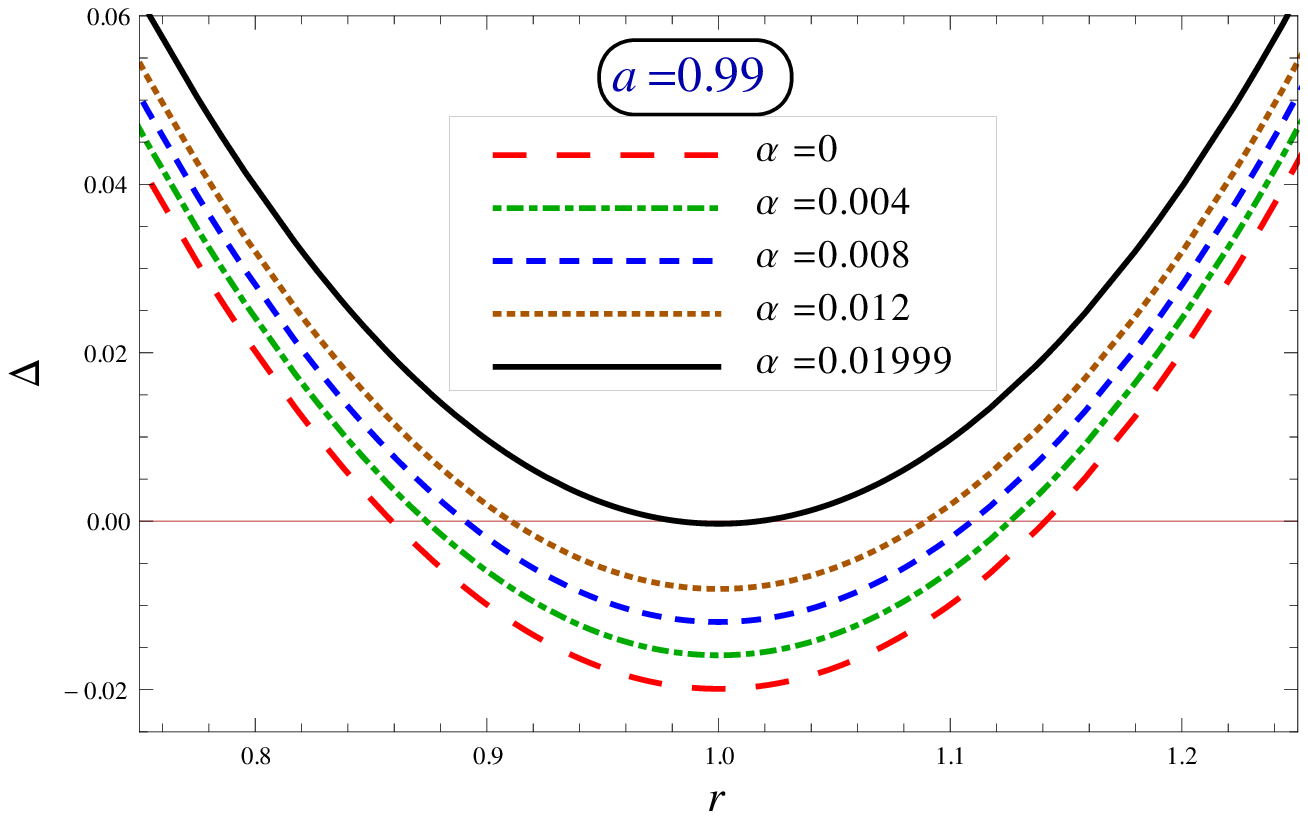}
 \end{tabular}
 \caption{These plots show the behavior of the event horizon $(\Delta=0)$ vs $r$ for different values of the deformation parameter $\alpha$. In these figures the solid (black) line corresponds to the extremal Kerr-MOG black hole and the large-dashed (red) line to the Kerr BH (i.e. when $\alpha=0$). All these plots are plotted for the value of the mass parameter $M_{\alpha}=1$.}\label{fig2_delta}
\end{figure*}

The behavior of the SLS is shown in Fig. \ref{fig1_gtt}, and that of the event horizon in Fig. \ref{fig2_delta}. It is clear from the figures that there exists a set of values for the parameters for which we have two horizons, i.e., an inner (Cauchy) horizon and an outer (event) horizon. In Fig. \ref{fig4_BH_NBH} and Table \ref{table1} we present the range of the parameter $\alpha$ for which a BH EH exists; from them one can easily conclude that the range of parameter $\alpha$ decreases as the value of spin parameter increases. Clearly, from Fig. \ref{fig5_ergo} the SLS and the EH surfaces touch each other at the poles ($\theta=0\text{, and}\, \theta=\pi$); otherwise the SLS is outside the horizon. The region between the SLS and EH is popularly known as the ergosphere and the reason for the name is that any massive object going into the ergosphere enables us to ``extract" energy from the spinning BH. One can also obtain the radius and the corresponding values of the parameters for which there exists an extremal BH (a BH for which both horizons coincide):
\begin{equation}
 \label{rext}
 r_{ext}=M_{\alpha}=|a|\sqrt{1+ \alpha} \quad  \mbox{\rm when} \quad M_{\alpha}^2 = a^2(1+ \alpha).
\end{equation}

\section{Geodesics of massive particles and photons in the Kerr-MOG BH}

\subsection{Geodesics of massive particles}

We start this section by studying the motion of a particle with mass $m=1$ orbiting in the background of a Kerr-MOG BH. The motion of stars around a BH, for instance, can be approximately described by these dynamics. The Hamilton-Jacobi equation guiding geodesic motion in this spacetime with the metric tensor $g^{ij}$ is given by
\begin{equation}
 \label{hj}
 2\frac{\partial S}{\partial \tau}=-g^{ij}\frac{\partial S}{\partial x^i}\frac{\partial S}{\partial x^j},
\end{equation}
where $S$ denotes Hamilton's principal function, and $\tau$ is an affine
parameter along the geodesics. For this BH background, the Hamilton's principal function $S$ can be separated as
\begin{equation}
 \label{hp}
 S=\frac{1}{2}\tau-Et+L\phi+S_{r}(r)+S_{\theta}(\theta),
\end{equation}
where $S_{r}$ and $S_{\theta}$ are functions of $r$ and $\theta$, respectively. The constants $E$ and $L$ correspond to conserved energy and angular momentum per unit mass through the normalization condition $1=-p_{i}p^{i}$ and are given by the following equations
\begin{eqnarray}
\label{E}
 E &=&\frac{\tilde{E}}{m}=-g_{ij}\xi^{i}U^{j}, \\
 \label{L}
  L &=&\frac{\tilde{L}}{m}=g_{ij}\psi^{i}U^{j}.
\end{eqnarray}
There is also a Carter constant of motion ${\cal K}$ due to the existence of a Killing tensor $K_{\mu\nu}$ for the Kerr-MOG metric. Its derivation a straightforward generalization of the Carter constant of motion for the Kerr spacetime \cite{Carter} in the sense that the only difference resides in the new form of the $\Delta$ function defined in (\ref{Delta}) compared to the same function for the Kerr BH metric (see \cite{HN_2015} as well for details). Thus, the Carter constant of motion defined according to the Hamilton-Jacobi method reads
\begin{eqnarray}
{\cal K} \equiv K_{ij}U^{i}U^{j} - (L-aE)^2 = C - (L-aE)^2,
\end{eqnarray}
where $C$ is a constant that arises from the contraction $K_{ij}U^{i}U^{j}$; the Carter constant provides a measure of how much the trajectory of a test particle departs from the equatorial plane $\theta = \pi/2$, where this quantity vanishes. Thus, the necessary and sufficient condition for a particle that initially is located in the equatorial plane to remain in it is to have ${\cal K}=0$. Any particle which crosses the equatorial plane necessarily has ${\cal K}>0$.

\begin{figure}
\includegraphics[scale=0.68]{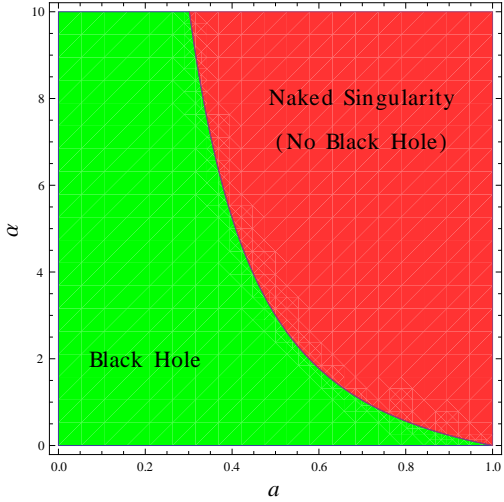}
 \caption{Plot showing the plane of the deformation parameter $(\alpha)$ vs the spin parameter $(a)$ of the Kerr-MOG BH for $M_{\alpha}=1$. The curve separates the BH from no BH regions (i.e., where there is no real root of $\Delta = 0$).}
\label{fig4_BH_NBH}
\end{figure}

Thus, the geodesics for massive particles in the Kerr-MOG BH background can be obtained as first-order differential equations for every direction following the procedure presented in \cite{Bardeen} and have the form
\begin{eqnarray}
\Delta\Sigma\,U^{t}\!=\!\left[(\!r^2\!+\!a^2)^2\!-\!\Delta a^2 \sin^2\!\theta \right]\!E\!-\!(r^2\!+\!a^2\!-\!\Delta)aL,   \label{Ut}
\end{eqnarray}
\begin{eqnarray}
\Sigma^2 (U^{r})^2\!=\!T^2\!-\!\Delta\left[r^2\!+\!(L\!-\!aE)^2\!+\!{\cal K}\right] \equiv V_r^2(r),
\label{V2}
\end{eqnarray}
\begin{equation}
\Sigma^2(U^{\theta})^2\!=\!{\cal K}\!-\!\left[a^2(1\!-\!E^2)\!+\!\frac{L^2}{\sin^2 \theta}\right]\cos^2\theta\!\equiv\!\Theta^2(\theta),
\label{Theta2}
\end{equation}
\begin{eqnarray}
\left(\!\Delta\Sigma\sin^2\!\theta\!\right)\!U^{\phi}\!\!=\!a\sin^2\!\theta(\!r^2\!+\!a^2\!\!-\!\Delta\!)E\!+
\!(\!\Delta\!-\!a^2\sin^2\!\theta)\!L,  \label{Uphi}
\end{eqnarray}
where $V_r^2(r)$ and $\Theta^2(\theta)$ are functions that respectively depend on the $r$ and $\theta$ coordinates alone, and we have introduced the following quantity
\begin{eqnarray}
 \label{T}
 T&\equiv&E\left(r^2+a^2\right)-La.
\end{eqnarray}
For bounded orbits we have $ E < 1$, while for unbounded orbits we obtain $E \geq 1\,$ (see \cite{Bardeen,Wilkins} for details).

Thus, the geodesic equations of a massive test particle moving in the Kerr-MOG BH background with given constants of motion $E$, $L$, ${\cal K}$, and initial conditions $x^{i}_{0}$ are encoded in $U^{i}$ and are given by the relations (\ref{Ut})--(\ref{Uphi}).

A similar calculation of the circular orbits of massive particles in the Kerr-MOG spacetime was performed in \cite{PLR}; moreover, in \cite{LR} the authors have also studied more general trajectories of massive particles moving around this spacetime metric.

\subsection{Geodesics of photons}

The same procedure can be performed to obtain the null geodesics of photons with 4-momentum $k^i$ moving outside the event horizon of the Kerr-MOG BH spacetime. In this case, the null character of the geodesics implies that $k_ik^i=0$ and the motion of photons preserve the following constants quantities
\begin{eqnarray}
  E_{\gamma} &=& - g_{ij} \xi^{i} k^{j}\,, \label{Egamma}  \\
  L_{\gamma} &=& g_{ij} \psi^{i} k^{j}\,, \label{Lgamma}\\
  Q_{\gamma} &\equiv& K_{ij}k^{i}k^{j}\!-\!(L_{\gamma}\!-\!aE_{\gamma})^2\!=\!C_{\gamma}\!-\!(L_{\gamma}\!-\!aE_{\gamma})^2,
\end{eqnarray}
where $C_{\gamma}$ is a constant. In \cite{CVE}, a detailed analysis of null geodesics in an arbitrary spacetime is given.

Therefore, the geodesic equations that describe the motion of photons in the Kerr-MOG BH spacetime with given constant parameters $E_{\gamma}$, $L_{\gamma}$, $Q_{\gamma}$, and initial conditions $y^{i}_{0}$ are parametrized by the components of the
4-momentum $k^{i}$ and read
\begin{eqnarray}
\Delta\Sigma k^{t}\!=\!
\left[(r^2\!+\!a^2)^2\!\!-\!\Delta a^2\!\sin^2\!\theta\right]\!E_{\gamma}\!-\!(r^2\!+\!a^2\!-\!\Delta)aL_{\gamma},
\end{eqnarray}
\begin{eqnarray}
\Sigma^2(k^{r})^2=T_{\gamma}^2-\Delta\!\left[\!(L_{\gamma}-aE_{\gamma})^2+Q_{\gamma}\right],
\label{kr}
\end{eqnarray}
\begin{eqnarray}
\Sigma^2(k^{\theta})^2 &=& Q_{\gamma} - \left[ - a^2 E_{\gamma}^2 + \frac{L_{\gamma}^2}{\sin^2 \theta} \right] \cos^2 \theta\,,
\label{ktheta}
\end{eqnarray}
\begin{eqnarray}
\left(\!\Delta\Sigma\sin^2\!\theta\!\right)\!k^{\phi}\!\!=\!a\sin^2\!\theta(r^2\!\!+\!a^2\!\!-\!\Delta)\!E_{\gamma}\!\!+\!\!(\Delta\!\!-\!a^2\!\sin^2\!\theta\!)\!L_{\gamma},
\end{eqnarray}
where again the right-hand side of the expressions (\ref{kr}) and  (\ref{ktheta}) depend only on the radial $r$ and polar $\theta$ coordinates, respectively, and we have defined
\begin{eqnarray}
 \label{Tgamma}
 T_{\gamma}=(r^2+a^2)E_{\gamma}-aL_{\gamma}.
\end{eqnarray}

Thus, at this point we have completely characterized the motion of massive particles (which can approximately describe the motion of stars) and photons in the gravitational field of the Kerr-MOG BH in terms of the 4-velocity $U^i$ and the 4-momentum $k^i$.

\subsection{Geodesics of massive particles in the equatorial plane}

Now we shall consider the motion of massive and massless particles in the equatorial plane $(\theta=\pi/2)$ as a particular case, implying that both $U^{\theta}$ and $k^{\theta}$ vanish and that ${\cal K}=0=Q_{\gamma}$.

Thus, in this case the components of the $4$-velocity adopt a simple form
 \begin{eqnarray}
 \label{tdot}
 r^2U^t&=&a\left(L-aE\right)+\left(r^2+a^2\right)\frac{T}{\Delta},\\
 \label{rdot}
 r^2U^r&=&\pm \sqrt{V_{r}},\\
  \label{pdot}
 r^2U^\phi&=& \left(L-aE\right)+\frac{aT}{\Delta}.
\end{eqnarray}
Here $V_{r}$ is an ``effective potential" governing the particle motion in the radial coordinate $r$.

We shall further consider circular trajectories, a relevant class of orbits that are very important to get physical insight about the dynamics of orbiting stars around a BH; however, for real applications of this method it is necessary to consider more general classes of stars' orbits, elliptical and not restricted to lie in the equatorial plane, in particular.

For circular orbits in the equatorial plane the radial component of the 4-velocity (\ref{rdot}) must vanish at a fixed distance $r$, i.e., $U^r=0$, and possess a minimum that allows for bound orbits. Thus, Eq. (\ref{rdot}) gives the condition on the effective potential and on its first derivative as \cite{Bardeen,Wilkins}
\begin{equation}
\label{Vmin}
V_{r}=0,   \qquad \mbox{ and} \qquad \frac{dV_{r}}{dr}=0.
\end{equation}
Remarkably, these equations can be nontrivially solved for the conserved energy $E$ and the conserved angular momentum $L$ to give
\begin{eqnarray}
 \label{E1}
 E&=&\frac{1}{\sqrt{Q_\pm}}\left[1-2\frac{M_\alpha}{r}+\frac{M_\beta}{r^2}\pm a \sqrt{\frac{\varpi}{r^3}}\right],\\
 \label{L1}
 L&=& \pm \sqrt{\frac{r\varpi}{Q_\pm}}\left[\frac{a^2}{r^2}+1\mp a\sqrt{\frac{1}{r^3\varpi}}\left(2M_\alpha-\frac{M_\beta}{r}\right)\right],\nonumber\\
\end{eqnarray}
where we have defined
\begin{eqnarray}
 M_\beta&=&\frac{\alpha}{\left(1+\alpha\right)}M_\alpha^2,\\
 Q_\pm&=&1+\frac{2M_\beta}{r^2}-\frac{3M_\alpha}{r}\pm 2a\sqrt{\frac{\varpi}{r^3}},\\
 \varpi&=& M_\alpha-\frac{M_\beta}{r}. \label{pi}
\end{eqnarray}
One can easily check that Eqs. (\ref{E1}) and (\ref{L1}) reduce to conserved energy $E$ and angular momentum $L$ for the Kerr BH as the parameter $\alpha\rightarrow 0$. In the Appendix we provide a detailed computation of these quantities following the algorithm of \cite{chandra}.

By substituting the expressions (\ref{E1}) and (\ref{L1}) into Eqs. (\ref{tdot}) and (\ref{pdot}), we obtain useful expressions for the 4-velocity components in terms of the metric parameters
\begin{eqnarray}
\label{tdot1}
 U^t&=&\frac{r^{\frac{3}{2}}\pm a\sqrt{\varpi}}{r^{\frac{3}{2}}\sqrt{Q_{\pm}}},\\
 \label{pdot1}
 U^{\phi}&=&\frac{\pm \sqrt{\varpi}}{r^{\frac{3}{2}}\sqrt{Q_{\pm}}}.
\end{eqnarray}

Finally, the condition for stable orbits is given by
\begin{equation}
 \label{ddveff}
 \frac{d^{2}V_{r}}{dr^2} = \left(6r^2+a^2\right)\left(E^2-1\right)-L^2+6M_{\alpha}r-M_{\beta}\le 0.
\end{equation}
This condition sets a bound that must be obeyed by the radial coordinate $r$ in order for the star to have a stable orbit:
\begin{equation}
 \label{stabilitycondn}
 M_{\alpha}r\Delta-4\left(M_{\alpha}r-M_{\beta}\right)\left(\sqrt{M_{\alpha}r-M_{\beta}}\mp a\right)^2\ge 0.
 \end{equation}
This relation leads to an algebraic expression of sixth order for $r$ that cannot be solved analytically.

\subsection{Geodesics of photons in the equatorial plane}

We now consider the null geodesics $k^{i}k_{i} =0$ of photons which move outside the event horizon of the modified Kerr BH along the equatorial plane where $\theta=\pi/2$.

Similarly, as we obtained Eqs. (\ref{tdot})-(\ref{pdot}) in terms of conserved energy $E$ and angular momentum $L$, we can also write the null geodesics of photons, parametrized by the 4-momentum components $k^{i}$, in the equatorial plane in terms of $E_{\gamma}$ and $L_{\gamma}$ in the following way
\begin{eqnarray}
 \label{tdotg}
 r^2k^t&=&a\left(L{\gamma}-aE_{\gamma}\right)+\left(r^2+a^2\right)\frac{T_{\gamma}}{\Delta},\\
 \label{rdotg}
 r^2k^r&=&\pm \sqrt{T_{\gamma}^{2}-\Delta\left(L_\gamma-aE_\gamma\right)^2},\\
  \label{pdotg}
 r^2k^\phi&=& \left(L{\gamma}-aE_{\gamma}\right)+\frac{aT_{\gamma}}{\Delta}.
\end{eqnarray}
Since the motion of the photons is not bound, the radial component of the 4-momentum is in general different from zero.

\section{Red- and blueshifts of photons in the modified Kerr background}

Now we shall calculate the red- and blueshifts that emitted photons by massive objects orbiting around the modified Kerr spacetime experience while traveling along null geodesics in the direction of an observer situated far away from the source. Here, it is worth mentioning that the following algorithm deals  with the problem on the basis of directly observable quantities: the red- and blueshifts of photons, which are coordinate independent in comparison with the tangential velocities, which are coordinate dependent.

\begin{figure*}
\begin{tabular}{c c c}
\includegraphics[scale=0.42]{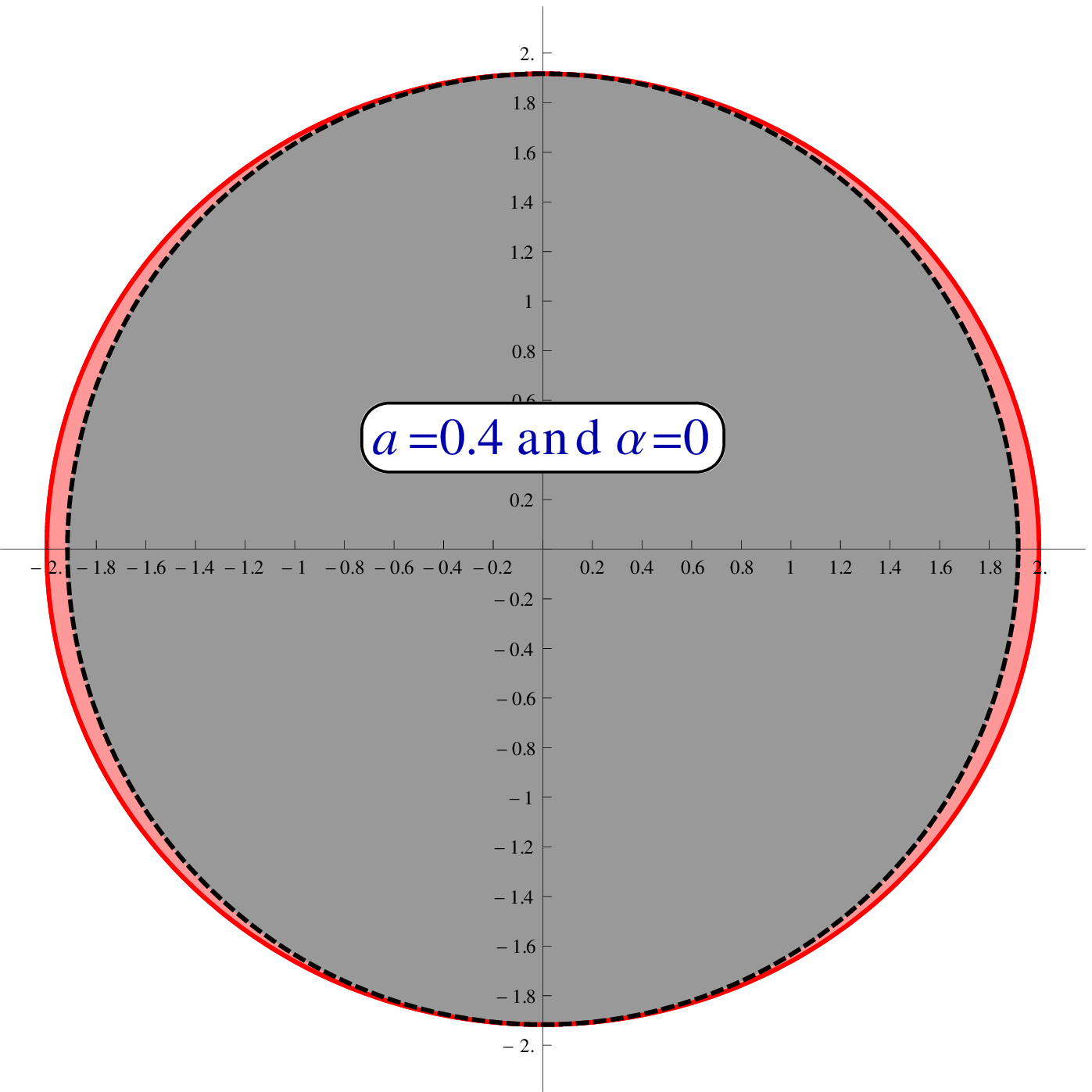}\hspace{0cm}
&\includegraphics[scale=0.42]{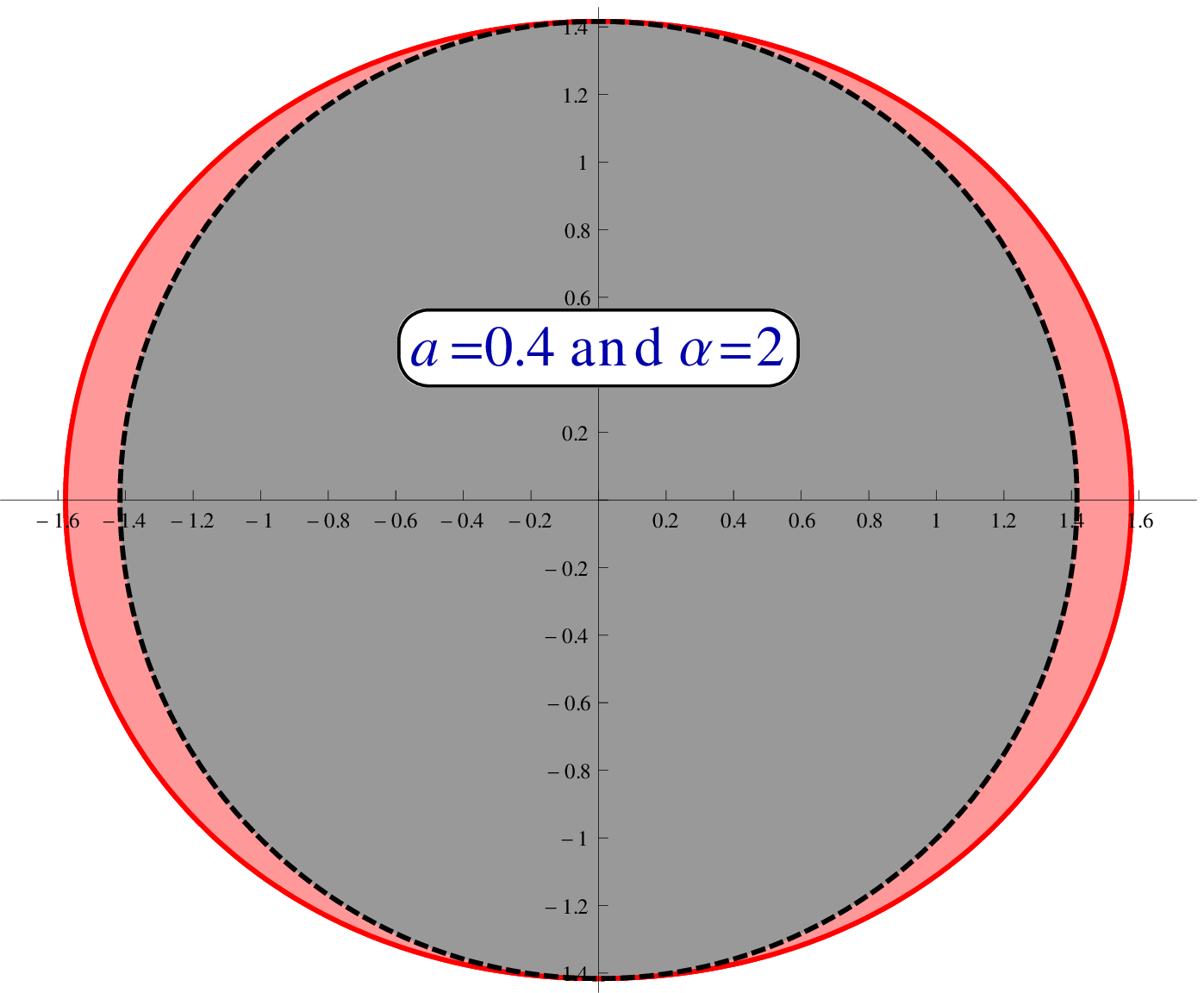}\hspace{0cm}
&\includegraphics[scale=0.42]{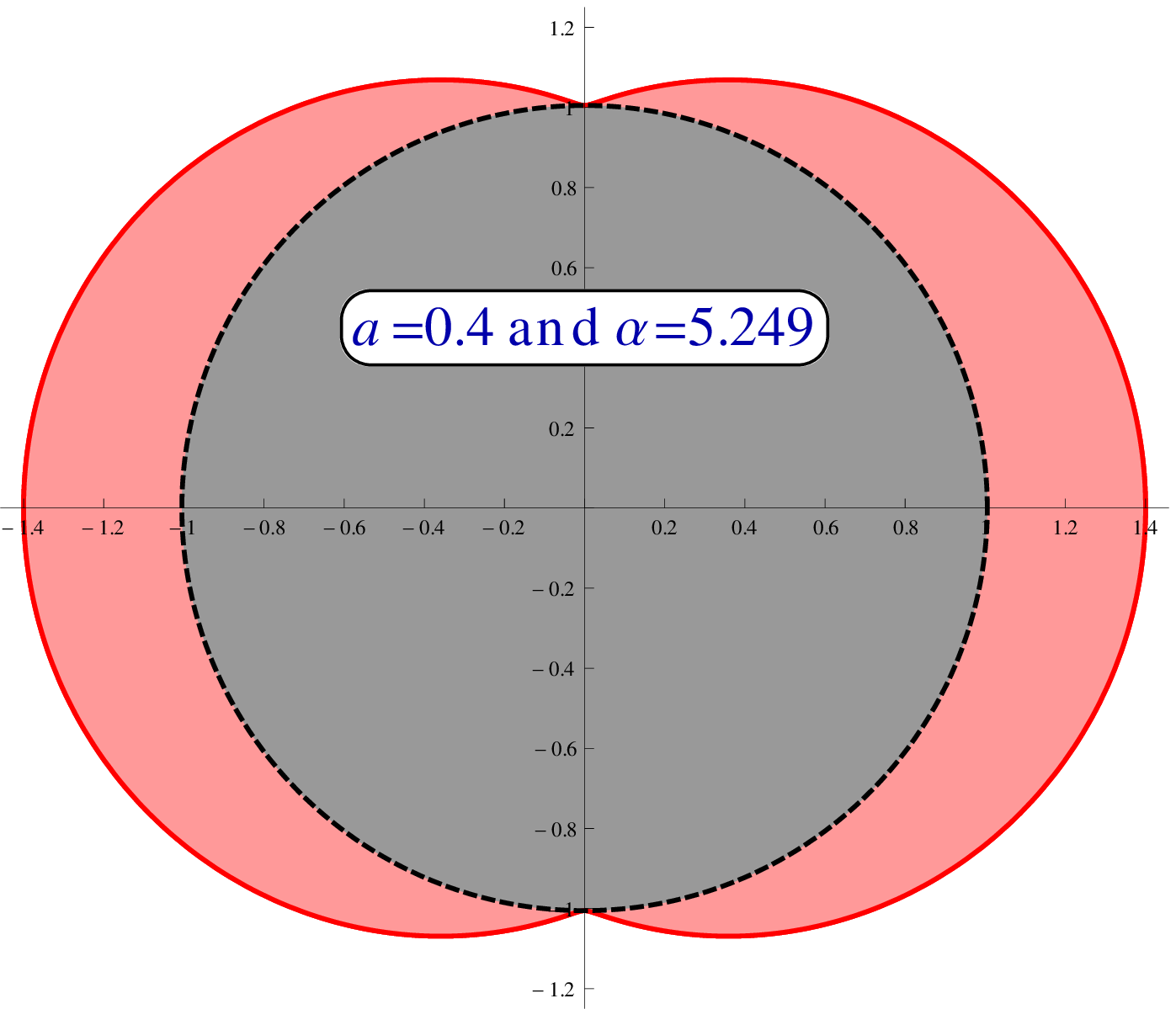}\hspace{0cm}
\\
\includegraphics[scale=0.42]{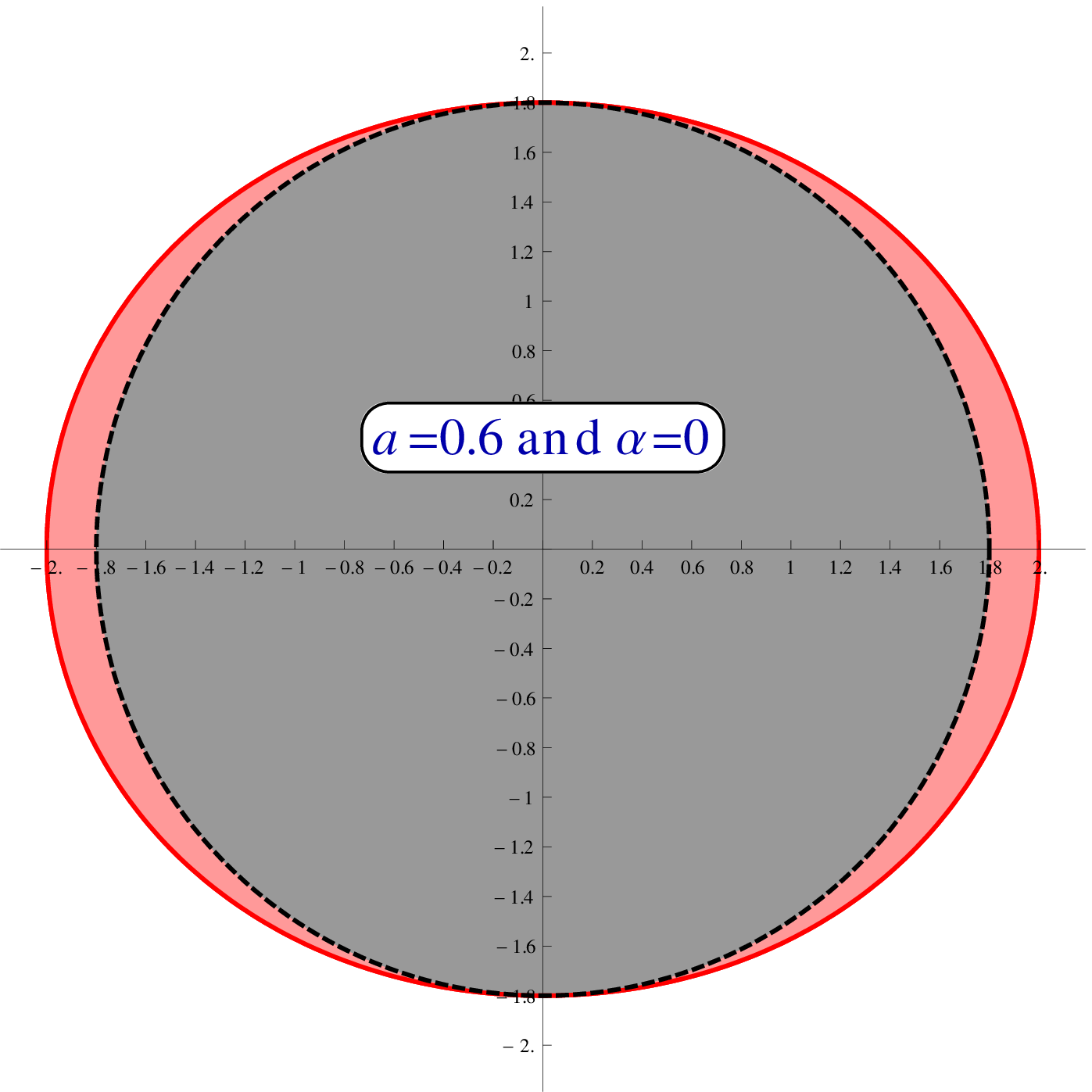}\hspace{0cm}
&\includegraphics[scale=0.42]{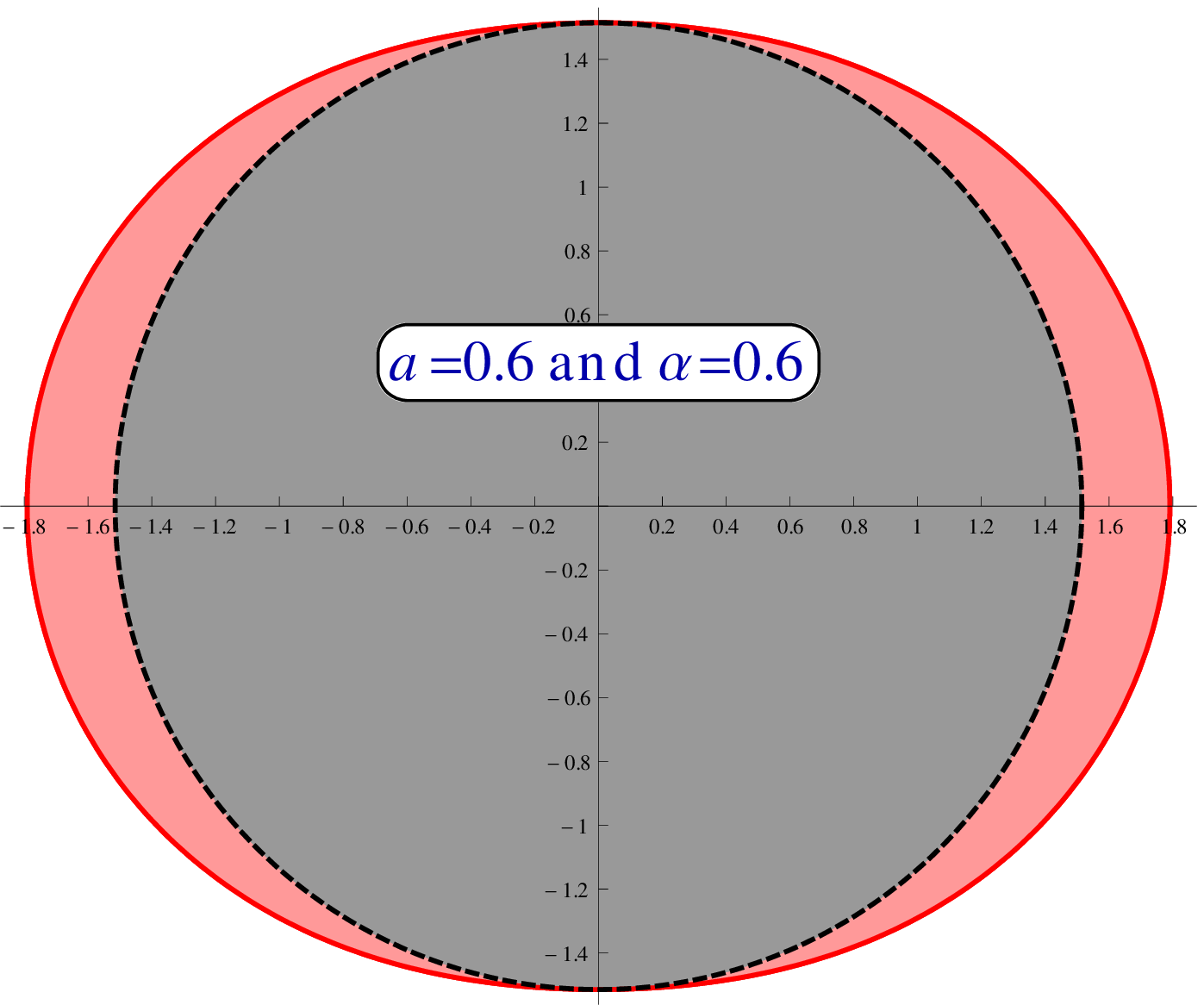}\hspace{0cm}
&\includegraphics[scale=0.42]{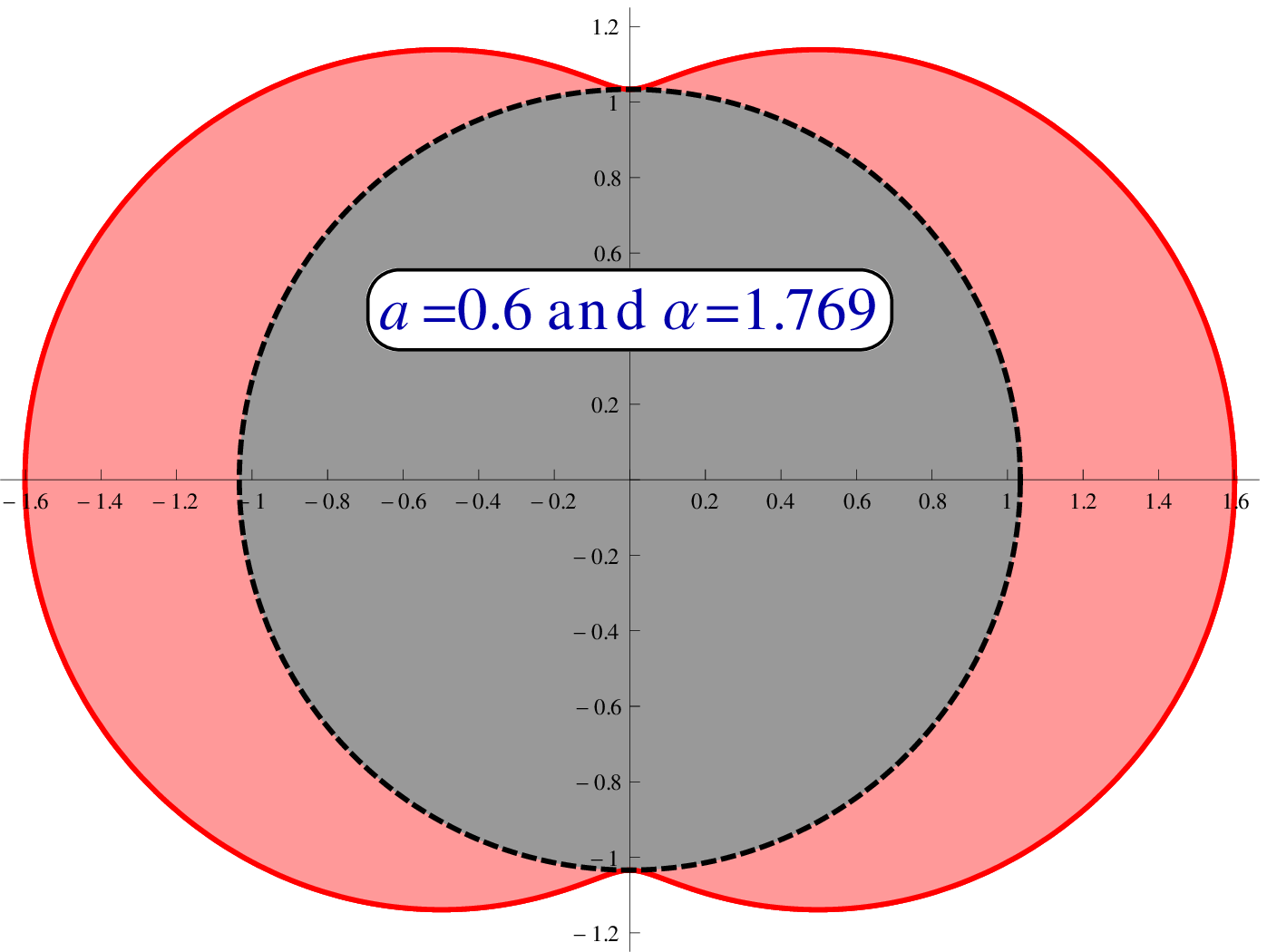}\hspace{0cm}
\\
\includegraphics[scale=0.42]{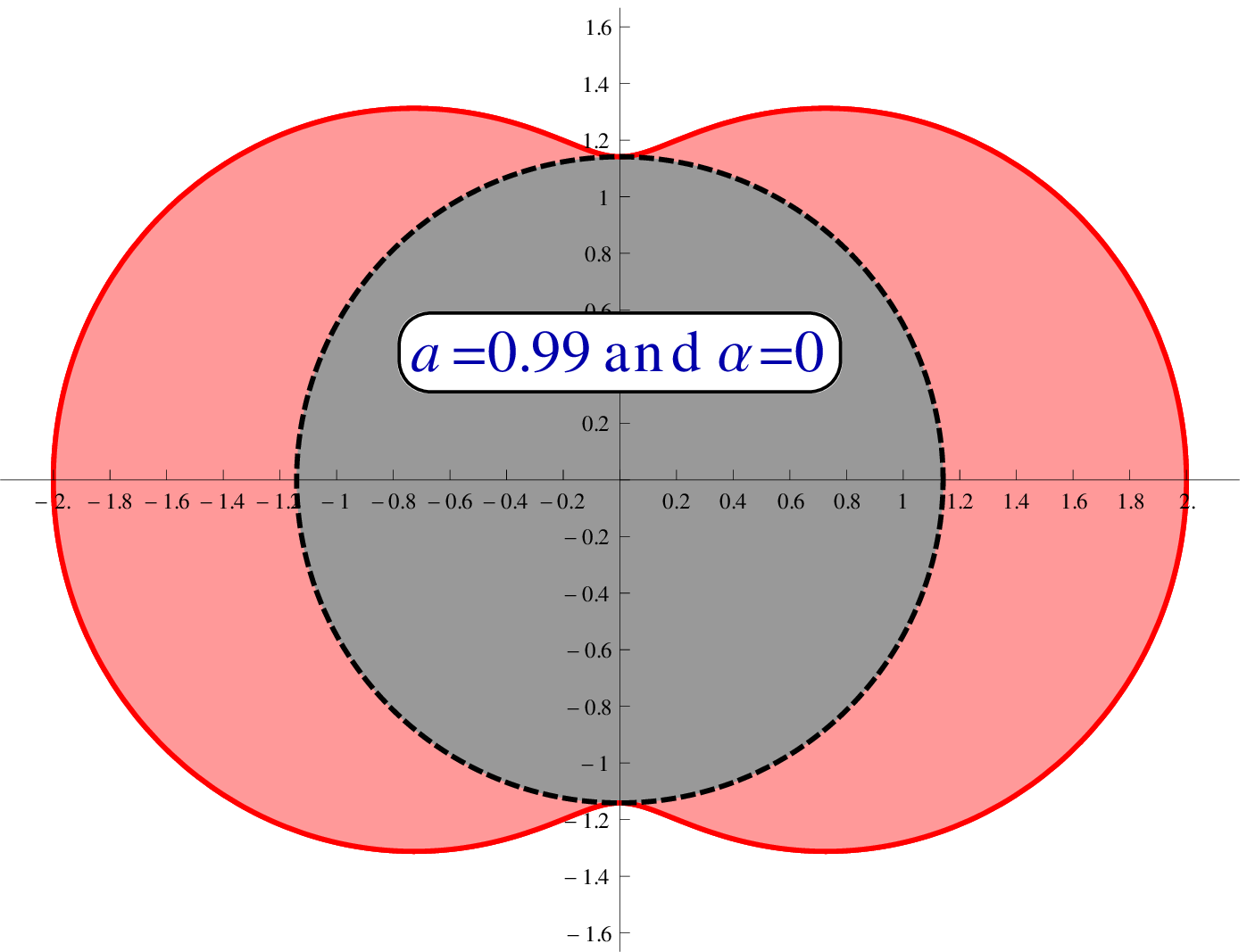}\hspace{0cm}
&\includegraphics[scale=0.42]{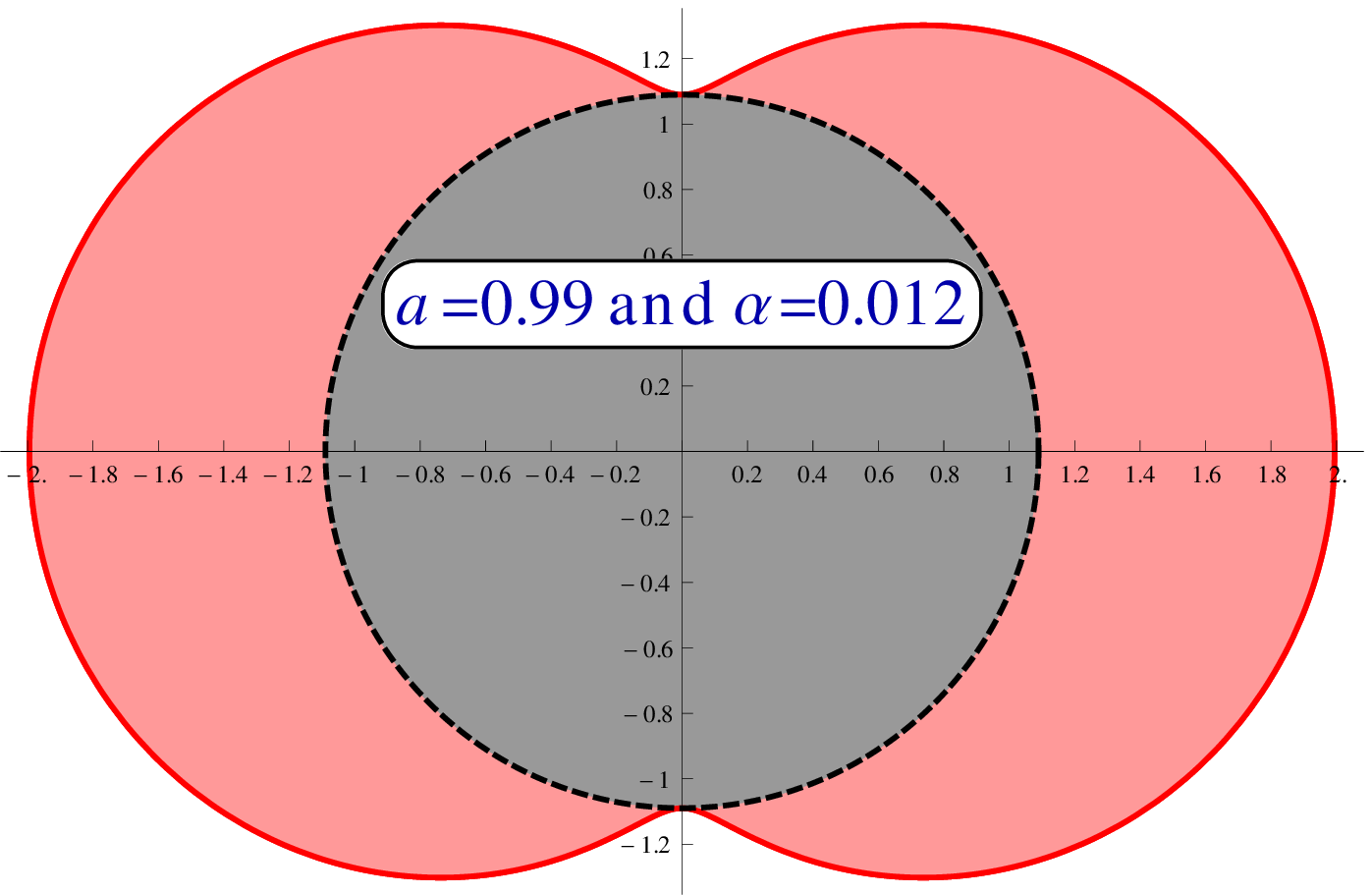}\hspace{0cm}
&\includegraphics[scale=0.42]{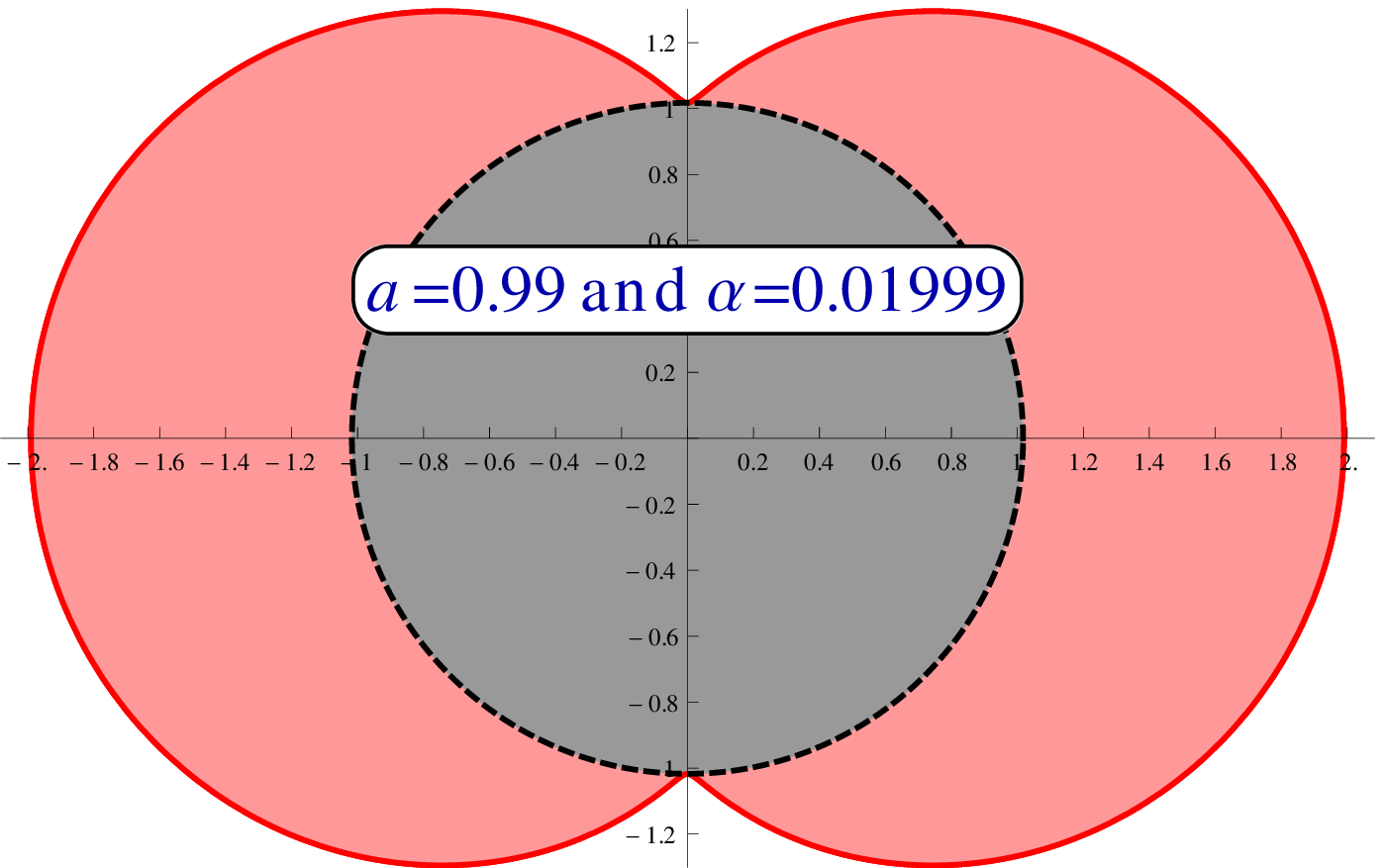}\hspace{0cm}
 \end{tabular}
 \caption{The shape of the ergosphere for the different values of the spin parameter $a$ is plotted. Here, the ergosphere region (pink region) increases with the increase in the value of the parameter $\alpha$ for a fixed value of the spin parameter $a$ [where, the solid (red) lines represent the SLS and the dashed (black) lines represent the EH]. Here, we keep the value of the mass parameter $M_{\alpha}=1$.}
 \label{fig5_ergo}
\end{figure*}


Now, we write the general expression for the frequency of a photon in terms of the 4-velocity $U^i_{e/d}$ and the 4-momentum $k^{i}_{e/d}$ measured by an observer located at point $D$,
\begin{equation}
\nu_{e/d}=-k_i U^i|_{D_{e/d}}\,,
\label{freq}
\end{equation}
where the index $e/d$ stands for emission $(e)$ or detection $(d)$ at the spacetime point $D$.

Therefore, the frequency of signals emitted by a comoving observer at the emission point $(e)$ and detected by an observer located far away from the source at point $(d)$, in terms of the 4-velocities of the emitter $U^i_e=(U^t,U^r,U^\theta,U^\phi)|_e$ and the detector
$U^i_d=(U^t,U^r,U^\theta,U^\phi)|_d$ is
\begin{eqnarray}
\label{freqed}
\nu_{e/d} &=& - (k_{i}U^{i})\mid_{e/d}\,,   \nonumber
\end{eqnarray}
where the $k_e^{i}\!=\!(k^t,k^r,k^\theta,k^\phi)|_e$ and $k_d^{i}\!=\!(k^t,k^r,k^\theta,k^\phi)|_d$ are the 4-momenta of photons at the emission and detection points, respectively.

The redshift of photons is defined as
\begin{eqnarray}
\label{z}
z &=& \frac{\nu_e - \nu_d}{\nu_d} =  \frac{\nu_e}{\nu_d} -1\,,
\end{eqnarray}
from where we can write the general expression for the redshift of photons with arbitrary motion in terms of the 4-velocity of the stars $U^i$ and the 4-momentum of the photons $k^{i}$ in the following form
\begin{eqnarray}
\label{redshift1}
1&+&z = \frac{\nu_e}{\nu_d} \nonumber  \\
&=& \frac{(Ek^{t} - Lk^{\phi} - g_{rr}U^{r}k^{r} - g_{\theta\theta}U^{\theta}k^{\theta})\mid_{e}}
{(Ek^{t} - Lk^{\phi} - g_{rr}U^{r}k^{r} - g_{\theta\theta}U^{\theta}k^{\theta})\mid_{d}}\nonumber \\
&=& \frac{(E_{\gamma}U^{t} - L_{\gamma}U^{\phi} - g_{rr}U^{r}k^{r} - g_{\theta\theta}U^{\theta}k^{\theta})\mid_{e}}
{(E_{\gamma}U^{t} - L_{\gamma}U^{\phi} - g_{rr}U^{r}k^{r} - g_{\theta\theta}U^{\theta}k^{\theta})\mid_{d}}\,,
\end{eqnarray}
where we have used Eqs. (\ref{E}), (\ref{L}), (\ref{Egamma}), and (\ref{Lgamma}) for the constants $E$, $L$, $E_\gamma$ and $L_\gamma$, respectively, and the definition for frequency (\ref{freq}), together with the expressions for the 4-velocity of the emitter and the detector, as well as for the 4-momentum of the emitted and the detected photons. Any of these formulas can be used for practical purposes.

\subsection{Redshifts in the equatorial plane}

For the special cases when the observer moves in the equatorial plane, $U^{\theta}_{e/d}$ vanishes; moreover, when the detector is {\it ideally} situated at an infinite distance from the source $(r \rightarrow \infty)$, we can conclude from Eqs. (\ref{tdot})-(\ref{pdot}) that both $U^{r}_{d}$ and ${U^{\phi}_d}$ vanish, while $U^t_d$ tends to $E=1$, rendering the following 4-velocity
    $U^{i}_{d} = (1, 0, 0, 0)\,$.     
%

Further, if we consider that photons are moving in the equatorial plane, then $k^\theta$ vanishes and the 4-momentum of the emitted or detected photon at point $D_{e/d}$ reads
\begin{equation}
\label{kgral}
   k_{e/d}^i=\left.\left(k^t,k^r,0,k^\phi\right)\right|_{e/d}.
\end{equation}
Thus, the general expression for the redshift in the case when both the photons and the stars are restricted to move along the equatorial plane is given by any of the following equations:
\begin{eqnarray}
\label{redshift1}
1+z &=& \frac{\nu_e}{\nu_d} \nonumber  \\
&=& \frac{(Ek^{t} - Lk^{\phi} - g_{rr}U^{r}k^{r})\mid_{e}}
{(Ek^{t} - Lk^{\phi} - g_{rr}U^{r}k^{r})\mid_{d}}\nonumber \\
&=& \frac{(E_{\gamma}U^{t} - L_{\gamma}U^{\phi} - g_{rr}U^{r}k^{r})\mid_{e}}
{(E_{\gamma}U^{t} - L_{\gamma}U^{\phi} - g_{rr}U^{r}k^{r})\mid_{d}}\,,
\end{eqnarray}
where we just have made use of Eq. (\ref{kgral}).

\subsection{Redshifts for circular equatorial orbits}

Here we shall return to circular orbits as a simple but important class of trajectories that allow us to get physical insight about the dynamics of the rotating stars around a BH as it was mentioned above. Thus, for circular orbits of stars (when $U^r=0$), the redshift (\ref{redshift1}) takes the  following simple form
\begin{equation}
\label{zcircorbits}
1+z =
\frac{\nu_e}{\nu_d} = \frac{\left.\left(E_\gamma U^t-L_\gamma
U^\phi\right)\right|_e}{\left.\left(E_\gamma U^t - L_\gamma U^\phi\right)\right|_d} =
\frac{U^t_e - b_e \,U^\phi_e}{U^t_d - b_d \,U^\phi_d}\,\,.
\end{equation}
This quantity can be used for the calculations of the mass and rotation (spin) parameter of the modified Kerr BH in terms of red- and blueshifts of photons ($z_{r}$  and $z_{b}$, see below) detected by a distant observer from the source; here, $b\equiv L_{\gamma}/E_{\gamma}$ has been defined as the apparent impact parameter.

We would like to point out here that corresponding to maximum and minimum values of the emitter frequency ($\nu_e$), there are two shifts, namely, the redshift ($z_r$, corresponding to a photon source which is going away from the observer) and the blueshift ($z_b$, corresponding to a photon source which is coming towards the observer), respectively. We shall further note that the conserved quantities $E_\gamma$ and $L\gamma$ are constant throughout (from the point of emission till the point of detection) along the null geodesics, hence $b_e=b_d$ should also remain constant throughout the photons' geodesics.

\subsection{Kinematical redshifts and light bending}

We are now going to calculate the kinematic redshifts $z_k$ of photons on both sides of the central value of the impact parameter (i.e., $b=0$) as it is important from the astronomers' point of view because they usually use a kinematic redshift to account for their data.

To obtain $z_k$ we need first to calculate the redshift of the photon emitted by an object located at the position where $b=0$,
\begin{equation}
\label{zatbnull}
1+z_c = \frac{U^t_e}{U^t_d}
\,\,,
\end{equation}
and then to subtract this expression from Eq. (\ref{zcircorbits}):
\begin{eqnarray}
\label{zkin}
&z_{\textrm{k}}&\equiv (1+z)-(1-z_c)= z-z_c,\nonumber\\
&=&\!\frac{(U^t-b U^\phi)|_e}{(U^t-b U^\phi)|_d}\!-\!\frac{U^t_e}{U^t_d}=\frac{U^t_e U^\phi_d b_d - U^t_d U^\phi_e b_e}
{U^t_d\left(U^t_d - b_d\,U^\phi_d\right)}\,.
\end{eqnarray}

In order to take into account the bending of light due to gravitational field around the massive astrophysical object, i.e., the Kerr-MOG BH,
we need to find the correlation between the impact parameter $b$ and the radius $r$ of the emitter's (or detectors') circular orbit, i.e., the mapping $b(r)$, as discussed in detail in \cite{HN_2015}.

\begin{figure}
\centering
\includegraphics[width=0.55\textwidth]{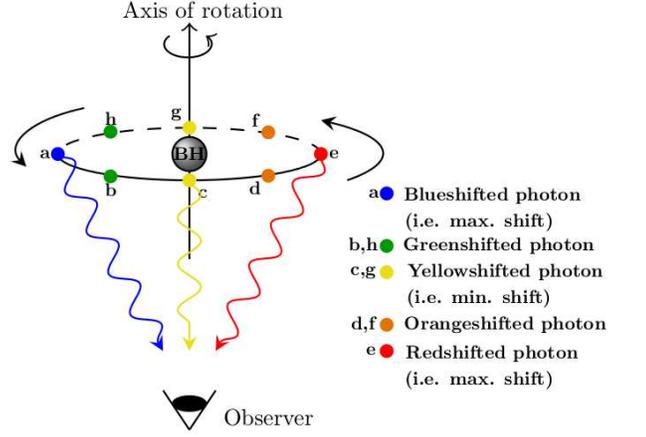}\hspace{0cm}
\caption{The ray diagram shows the three required measurements along the stars' trajectories revolving around the Kerr-MOG BH in the  equatorial plane in order to determine the kinematical red- and blueshifts of photons: two of them at the points where the impact parameter $b$ achieves its maximal value (points $a$ and $e$ where photons will be respectively blue- and redshifted for a corotating star), and another one where the impact parameter vanishes $b=0$ (point $c$ where the photons achieve their minimal shift).}
\label{fig5_zrzbz0}
\end{figure}

Hence, the maximized impact parameter $b$ for equatorial orbits is obtained from the null geodesic relation $k^ik_i=0$ by taking into account that $k^r=0$ and reads
\begin{equation}
\label{b}
b_\pm =  \frac{\left(-2M_\alpha r+M_\beta\right)a\pm r^2\sqrt{r^2+a^2-2M_\alpha r+M_\beta}}{r^2-2M_\alpha r+M_\beta} \,,
\end{equation}
where the two obtained values of the impact parameter $b_\pm$ can be either calculated at the emitter or detector position (recall that $b_e=b_d$ along the whole photons' trajectory). Furthermore, these two values give rise to two different shifts, namely $z_1$ and $z_2$ as pointed out in \cite{HN_2015}, which corresponds to the redshift of photons of a receding and an approaching object with respect to a distant observer:
\begin{equation}
\label{z+}
z_1 = \frac{U^t_e U^\phi_d\, b_{d_-} - U^t_d U^\phi_e\, b_{e_-}}{U^t_d\left(U^t_d - U^\phi_d\, b_{d_-}\right)}\,,
\end{equation}
\begin{equation}
\label{z-}
z_2 = \frac{U^t_e U^\phi_d\, b_{d_+} - U^t_d U^\phi_e\, b_{e_+}}{U^t_d\left(U^t_d - U^\phi_d\, b_{d_+}\right)}\,.
\end{equation}
Thus, the kinematical red- and blueshifts of photons require three measurements along the stars' trajectories to be determined: two of them at the points where the impact parameter $b$ achieves its maximal value, and another one where the impact parameter vanishes $b=0$ (see Fig. \ref{fig5_zrzbz0} for an illustrative picture).

We now define the angular velocity of an emitter or a detector which is situated at some distance from the source of photons in the following form
\begin{equation}
\label{omegaD}
\frac{U_{e/d}^\phi}{U_{e/d}^t}=\frac{d\phi}{dt}\equiv \omega_{e/d},
\end{equation}
where the subscripts $_e$ and $_d$ correspond to the angular velocity of the emitter and the detector of photons, respectively.

In terms of $\omega_d$, the shifts $z_1$ and $z_2$ can be written as
\begin{equation}
\label{z++}
z_1 = \frac{U^t_e\, \omega_d\, b_{d_-} - U^\phi_e\, b_{e_-}}{U^t_d\left(1 - \omega_d\, b_{d_-}\right)}\,,
\end{equation}
\begin{equation}
\label{z--}
z_2 = \frac{U^t_e\, \omega_d\, b_{d_+} - U^\phi_e\, b_{e_+}}{U^t_d\left(1 - \omega_d\, b_{d_+}\right)}\,.
\end{equation}
By knowing the expression for $U^{t}$ and $U^{\phi}$ from Eqs. (\ref{tdot1}) and (\ref{pdot1}), one can express the angular velocity of an object (i.e., photon source) revolving in a circular equatorial orbit around the modified-Kerr BH:
\begin{equation}
\label{omega}
\omega_{\pm}=\frac{\pm \sqrt{\varpi}}{r^{3/2} \pm a \sqrt{\varpi}}\,.
\end{equation}
Here, $+$ and $-$ signs in the angular velocity represent a corotating and a counterrotating object, respectively.

From Eq. (\ref{omegaD}) it is easy to see that Eq. (\ref{omega}) with the subscripts $_e$ and $_d$ corresponds to the angular velocity of the emitter and the detector of photons, respectively.

Now we can write the red- and blueshifts (\ref{z++}) and (\ref{z--}) by using Eqs. (\ref{tdot1}), (\ref{pdot1}), and (\ref{omegaD}) in the form
\begin{eqnarray}
\label{ZrKerr}
z_{r}&=&\frac{\sqrt{\varpi_e}r^{\frac{3}{2}}_{d}\sqrt{Q_{d\pm}}\,\omega_{d\pm}\left(\omega_{d\pm}b_{d_{-}}-\omega_{e\pm}b_{e_{-}}\right)}{\sqrt{\varpi_d}r^{\frac{3}{2}}_{d}\sqrt{Q_{e\pm}}\,\omega_{e\pm}\left(1-\omega_{d\pm}b_{d_{-}}\right)}\,,
\end{eqnarray}
\begin{eqnarray}
z_{b}&=&\frac{\sqrt{\varpi_e}r^{\frac{3}{2}}_{d}\sqrt{Q_{d\pm}}\,\omega_{d\pm}\left(\omega_{d\pm}b_{d_{+}}-\omega_{e\pm}b_{e_{+}}\right)}{\sqrt{\varpi_d}r^{\frac{3}{2}}_{d}\sqrt{Q_{e\pm}}\,\omega_{e\pm}\left(1-\omega_{d\pm}b_{d_{+}}\right)}\,,
\label{ZbKerr}
\end{eqnarray}
where, the subscripts $ _{\pm}$ stand for the corotating and counterrotating objects with respect to the direction of the angular velocity of the Kerr-MOG BH.

\section{Bounds on the redshifts of photons and a Kerr hypothesis test}

The above expressions for the red- and blueshifts of photons in terms of the Kerr-MOG BH parameters, mass $M_\alpha$, deformation parameter $\alpha$, and spin parameter $a$, as well as in terms of the detector radius $r_d$ and the radii of the orbits of stars (sources) $r_e$, are found to be
\begin{eqnarray}
\label{ZrK}
z_{r}&=&\pm\frac{r^{\frac{3}{2}}_{d}\sqrt{Q_{d\pm}}\left(2 M_{\alpha} r_{e}a-M_{\beta}a+r^{2}_{e}\sqrt{\Delta_{e}}\right)}{r^{\frac{3}{2}}_{e}\sqrt{Q_{e\pm}}\left(r^{\frac{3}{2}}_{d}\pm a\sqrt{\varpi_{d}}\right)}\times\nonumber\\
&&\frac{\left(r^{\frac{3}{2}}_{d}\sqrt{\varpi_e}-r^{\frac{3}{2}}_{e}\sqrt{\varpi_d}\right)}{r^{\frac{3}{2}}_{d}\left(\Delta_{e}-a^{2}\right)\pm r^{2}_{e}\sqrt{\varpi_{d}}\left(a+\sqrt{\Delta_{e}}\right)}\,,
\end{eqnarray}
\begin{eqnarray}
\label{ZbK}
z_{b}&=&\pm\frac{r^{\frac{3}{2}}_{d}\sqrt{Q_{d\pm}}\left(2 M_{\alpha} r_{e}a-M_{\beta}a-r^{2}_{e}\sqrt{\Delta_{e}}\right)}{r^{\frac{3}{2}}_{e}\sqrt{Q_{e\pm}}\left(r^{\frac{3}{2}}_{d}\pm a\sqrt{\varpi_{d}}\right)}\times\nonumber\\
&&\frac{\left(r^{\frac{3}{2}}_{d}\sqrt{\varpi_e}-r^{\frac{3}{2}}_{e}\sqrt{\varpi_d}\right)}{r^{\frac{3}{2}}_{d}\left(\Delta_{e}-a^{2}\right)\pm r^{2}_{e}\sqrt{\varpi_{d}}\left(a-\sqrt{\Delta_{e}}\right)}\,,
\end{eqnarray}
where we have made use of the relation $b_e=b_d$.

When $r_d>>M\ge a$ and the source is located at a far distance from the detector, the above Eqs. (\ref{ZrK}) and (\ref{ZbK}) reduce to
\begin{eqnarray}
\label{ZrKstatic}
z_{r} = \frac{\pm \sqrt{\varpi_e}\left(2M_{\alpha}r_{e}a-M_{\beta} a+r^{2}_{e}\sqrt{\Delta_{e}}\right)}
{r_e^{\frac{3}{2}}\,\sqrt{Q_{e\pm}}\left(\Delta_{e}-a^{2}\right)\,},
\end{eqnarray}
\begin{eqnarray}
z_{b} = \frac{\pm \sqrt{\varpi_e}\left(2M_{\alpha}r_{e}a-M_{\beta} a-r^{2}_{e}\sqrt{\Delta_{e}}\right)}
{r_e^{\frac{3}{2}}\,\sqrt{Q_{e\pm}}\left(\Delta_{e}-a^{2}\right)\,}.
\label{ZbKstatic}
\end{eqnarray}
These red- and blueshifts possess the following bounds with respect to the static ($a=0$) and extremal rotating ($|a|=M_{\alpha}/\sqrt{1+\alpha}$) limits of the Kerr-MOG BH:
\begin{eqnarray}
\label{Zrbounds}
z_{r}^{_{min}}\le z_{r} \le z_{r}^{_{max}} \qquad \textrm{and} \qquad  z_{b}^{_{min}}\le z_{b} \le z_{b}^{_{max}},
\end{eqnarray}
where we have defined
\begin{eqnarray}
\label{zrmin}
z_{r}^{_{min}}\!=\!\frac{\pm \sqrt{M_{\alpha}r_e-M_{\beta}}\,\,r_e}
{\sqrt{\left(r_e^2-3M_{\alpha}r_e+2M_{\beta}\right)\left(r_e^2-2M_{\alpha}r_e+M_{\beta}\right)}},
\nonumber
\end{eqnarray}
\begin{eqnarray}
\label{zrmax}
z_{r}^{_{max}}\!\!=\!\!\frac{\pm \sqrt{\varpi_e r_e}\left[\left(2M_{\alpha}r_e\!-\!M_{\beta}\right)M_{\alpha}\!+\!\sqrt{1\!+\!\alpha}r_e^2(r_e\!-\!M_{\alpha})\right]\!}
{r_e(\!\Delta_{e}\!\!-\!a^{2}\!)\!\sqrt{\!(\!1\!+\!\alpha\!)\!\left(\!r_e^2\!\!-\!\!3M_{\alpha}r_e\!\!+\!\!2M_{\beta}\!\right)\!\pm\!2M_{\alpha}\!\sqrt{\!(\!1\!+\!\alpha\!)\varpi_e r_e}}},
\nonumber
\end{eqnarray}
\begin{eqnarray}
\label{zbmin}
z_{b}^{_{min}}\!=\!\frac{\mp \sqrt{M_{\alpha}r_e-M_{\beta}}\,\,r_e}
{\sqrt{\left(r_e^2-3M_{\alpha}r_e+2M_{\beta}\right)\left(r_e^2-2M_{\alpha}r_e+M_{\beta}\right)}},
\nonumber
\end{eqnarray}
\begin{eqnarray}
\label{zbmax}
z_{b}^{_{max}}\!\!=\!\!\frac{\pm \sqrt{\varpi_e r_e}\left[\left(2M_{\alpha}r_e\!-\!M_{\beta}\right)M_{\alpha}\!-\!\sqrt{1\!+\!\alpha}r_e^2(r_e\!-\!M_{\alpha})\right]\!}
{r_e(\!\Delta_{e}\!\!-\!a^{2}\!)\!\sqrt{\!(\!1\!+\!\alpha\!)\!\left(\!r_e^2\!\!-\!\!3M_{\alpha}r_e\!\!+\!\!2M_{\beta}\!\right)\!\pm\!2M_{\alpha}\!\sqrt{\!(\!1\!+\!\alpha\!)\varpi_e r_e}}};
\nonumber
\end{eqnarray}
note that $z_{r}^{_{min}}$ and $z_{b}^{_{min}}$ have the same magnitude but different sign as it is expected for red- and blueshifts generated by the gravitational field of a nonrotating BH.

In the case when the deformation parameter $\alpha$ vanishes, we obtain the corresponding bounds on the red- and blueshifts of the Kerr BH metric:
\begin{eqnarray}
\label{ZrboundsKerr}
Z_{r}^{_{min}}\!\le z_{r}^{_{Kerr}}\!\le Z_{r}^{_{max}} \quad \textrm{and} \quad  Z_{b}^{_{min}}\!\le z_{b}^{_{Kerr}}\!\le Z_{b}^{_{max}}\!,
\end{eqnarray}
where now we have introduced
\begin{eqnarray}
\label{zrminKerr}
Z_{r}^{_{min}}\!=\!\frac{\pm \sqrt{M r_e}\,}
{\sqrt{\left(r_e-3M \right)\left(r_e-2M\right)}},
\nonumber
\end{eqnarray}
\begin{eqnarray}
\label{zrmaxKerr}
Z_{r}^{_{max}}=\frac{\pm \sqrt{M}\left[2M^2+r_e(r_e-M)\right]}
{(r_e-2M)\sqrt{r_e\left(r_e^2-3Mr_e\pm2M\sqrt{Mr_e}\right)}}\,,
\nonumber
\end{eqnarray}
\begin{eqnarray}
\label{zbminKerr}
Z_{b}^{_{min}}\!=\!\frac{\mp \sqrt{M r_e}\,}
{\sqrt{\left(r_e-3M \right)\left(r_e-2M\right)}},
\nonumber
\end{eqnarray}
\begin{eqnarray}
\label{zbmaxKerr}
Z_{b}^{_{max}}=\frac{\pm \sqrt{M}\left[2M^2-r_e(r_e-M)\right]}
{(r_e-2M)\sqrt{r_e\left(r_e^2-3Mr_e\pm2M\sqrt{Mr_e}\right)}}\,.
\nonumber
\end{eqnarray}

\begin{figure*}
\begin{tabular}{c c}
\includegraphics[scale=0.55]{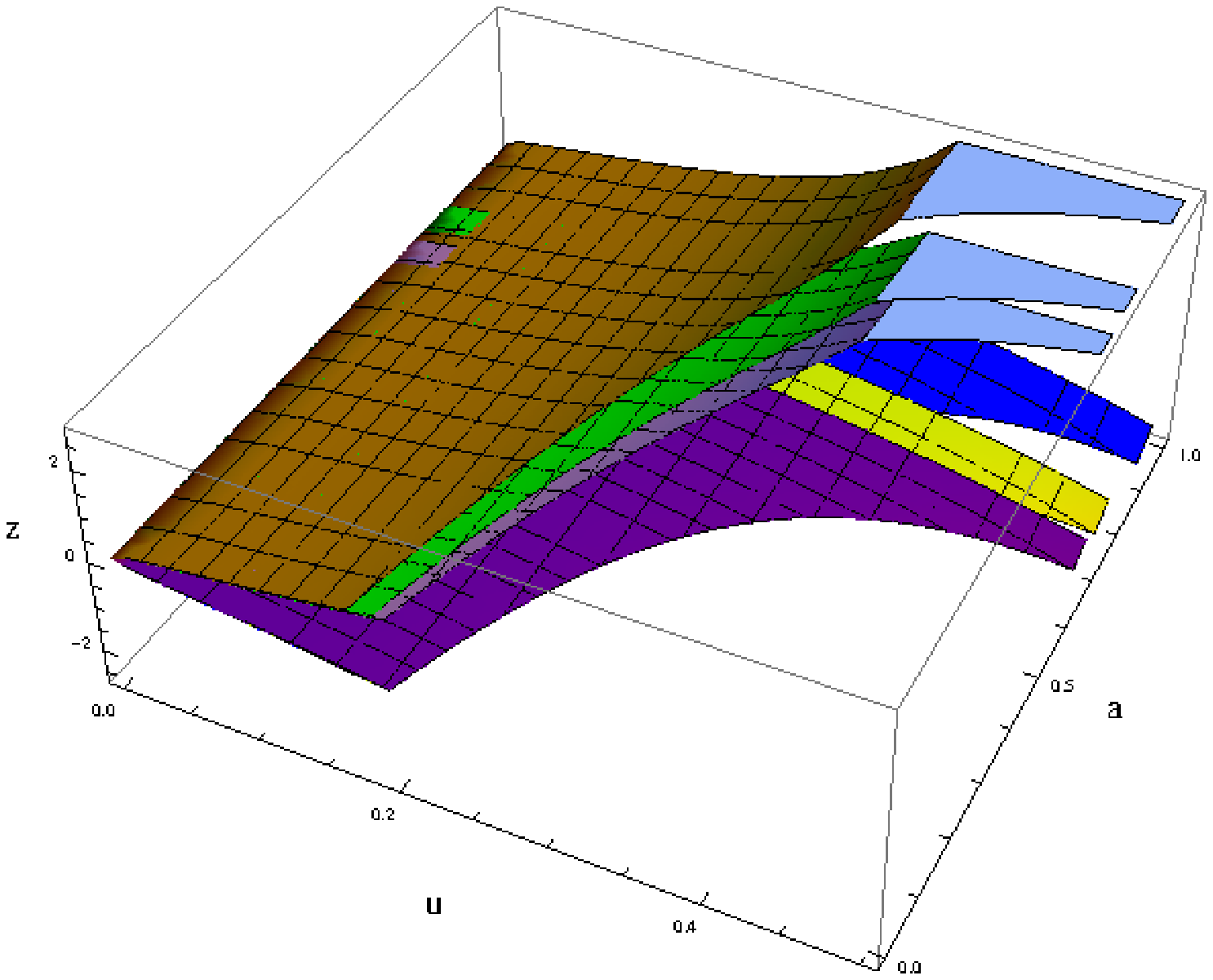}\hspace{0cm}
&\includegraphics[scale=0.55]{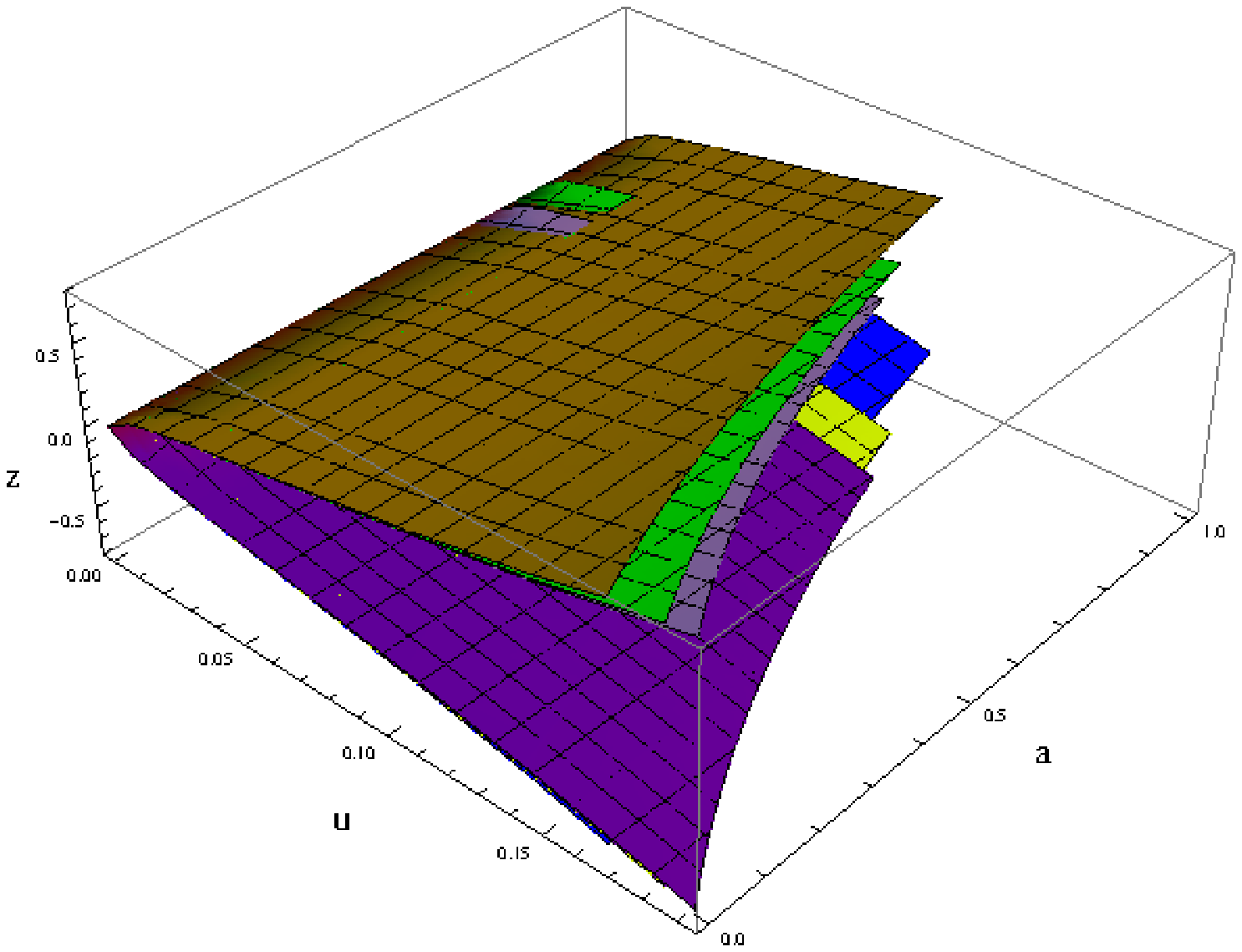}\hspace{0cm}
 \end{tabular}
\caption{The left (right) panel plot shows the shifts (named $z$ in the vertical axis) of photons emitted by corotating (counterrotating) particles around the Kerr-MOG black hole vs the rotation parameter $a$ and the variable $u = M_{\alpha}/r_e$. The brown (redshift) and blue (blueshift) surfaces correspond to the deformation parameter $\alpha=0$ (Kerr BH case), the green (redshift) and yellow (blueshift) surfaces correspond to $\alpha=0.5$, and the violet (redshift) and purple (blueshift) surfaces correspond to $\alpha=0.9$.}
\label{fig6_Shifts}	
\end{figure*}

\begin{figure*}
\begin{tabular}{c c}
\includegraphics[scale=0.55]{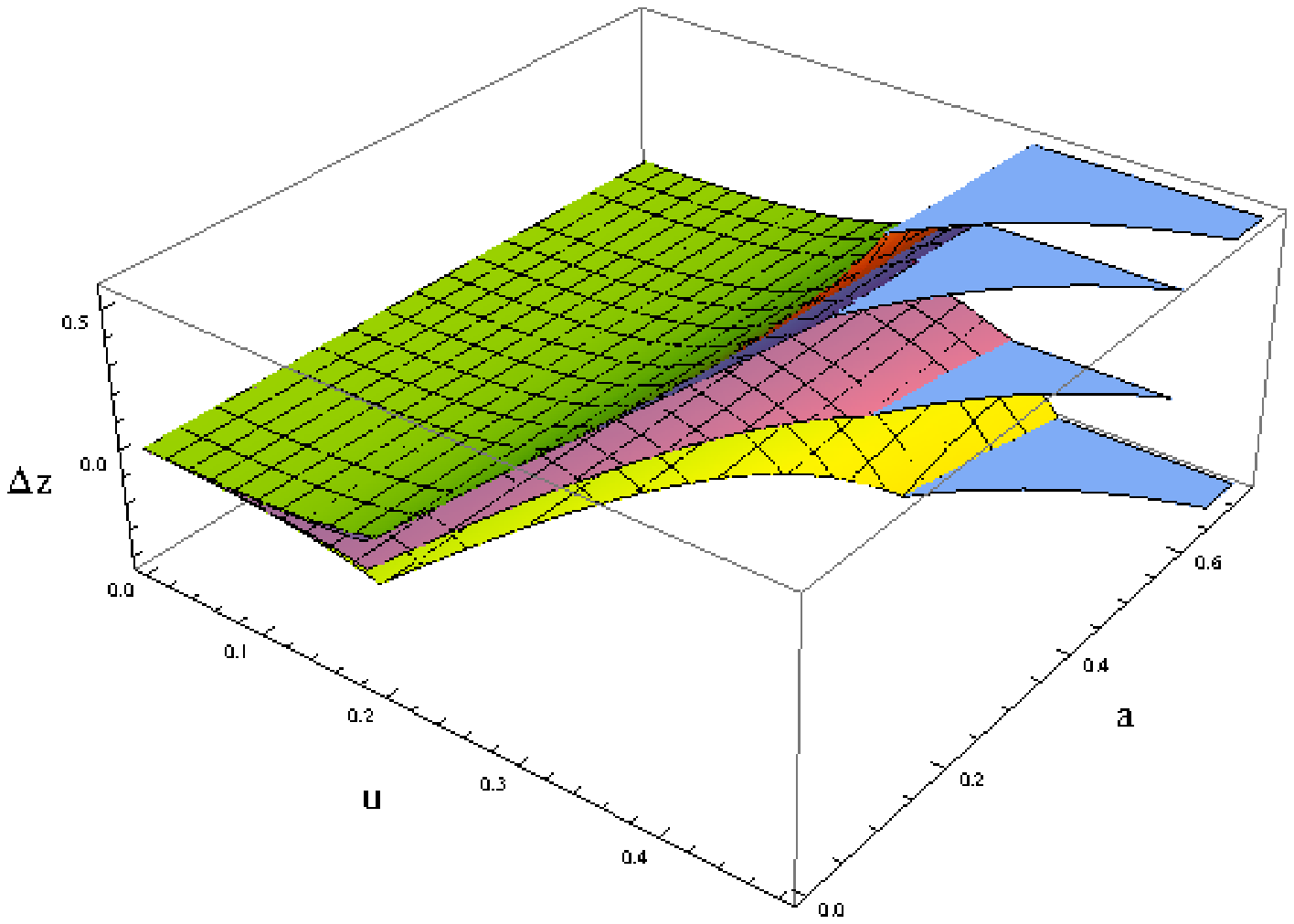}\hspace{0cm}
&\includegraphics[scale=0.55]{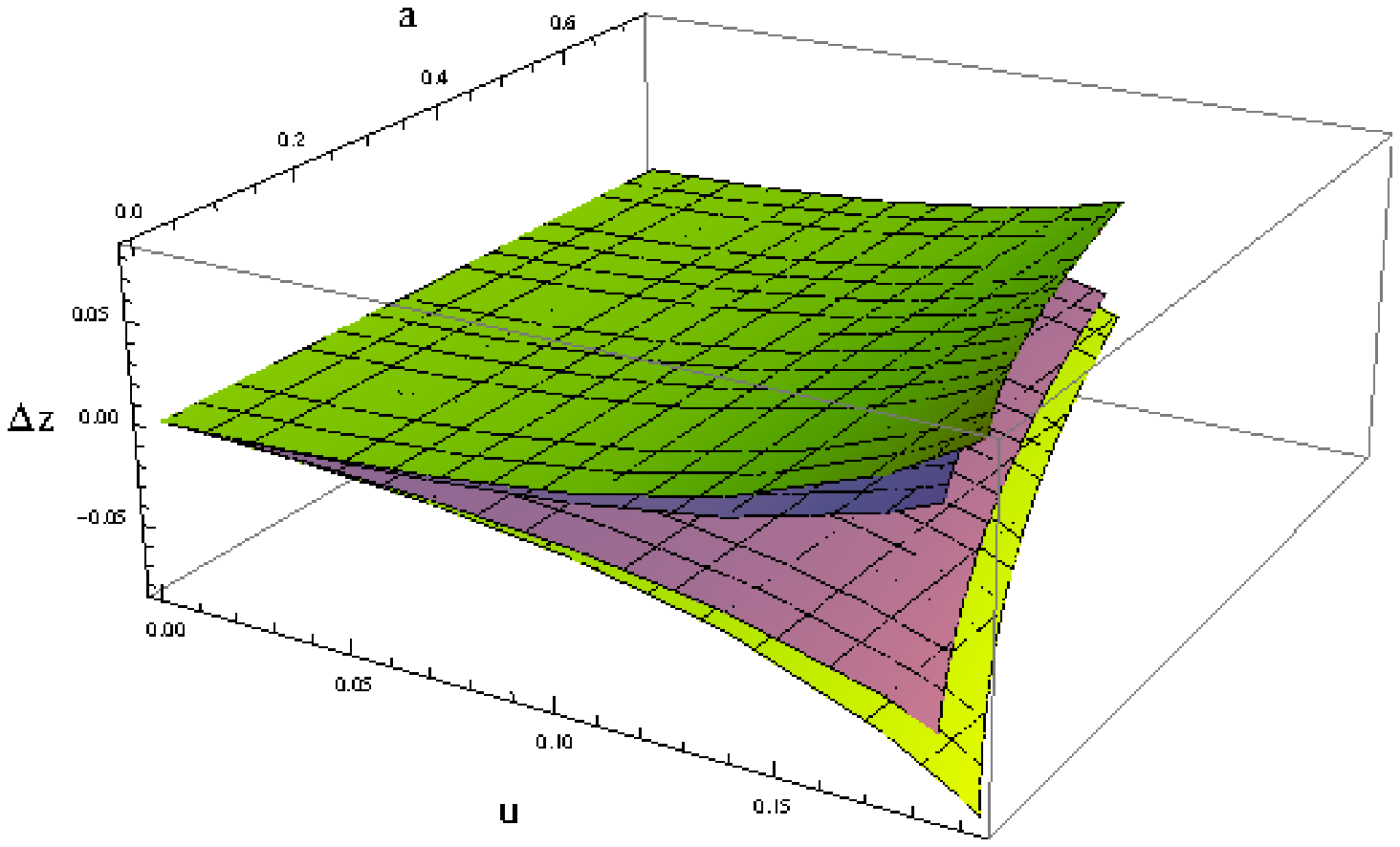}
\end{tabular}
\caption{The left (right) panel plot shows the differences in shifts of photons emitted by corotating (counterrotating) particles around the Kerr-MOG black hole vs the rotation parameter $a$ and the variable $u = M_{\alpha}/r_e$. The purple/violet color corresponds to the difference $\Delta z = z(\alpha=0)-z(\alpha=0.5)$ and the green/yellow one to $\Delta z = z(\alpha=0)-z(\alpha=0.9)$. }
\label{fig7_DShifts}
\end{figure*}

Thus, if a given set of observational data of red- and blueshifts falls within the intervals (\ref{ZrboundsKerr}), this implies a confirmation of the
Kerr black hole hypothesis in the strong gravitational regime and this would also impose some observational constraints on the deformation parameter of the Kerr-MOG BH and on the parameters that have the Kerr metric as a particular BH solution.

On the other hand, if the set of data does not fall within the intervals (\ref{ZrboundsKerr}) predicted by the GTR for the Kerr BH metric, but are allowed by the bounds (\ref{Zrbounds}) corresponding to the Kerr-MOG BH, we would have a breakdown of the GTR in favor of the STVG theory and we should have to describe the observed star dynamics with the latter theory or any theory whose parameters allow for such set of observations.

It could also happen that the observed red- and blueshifts do not match the predictions made by any one of the Kerr or Kerr-MOG BH metrics, implying that we shall need to look for an alternative theory of gravity in order to consistently explain the observational data.

The bounded red- and blueshifts of photons for different values of the deformation parameter $\alpha$ (for both corotating and counterrotating cases), including the Kerr BH case for comparison, are plotted in Fig. \ref{fig6_Shifts} as a function of the rotation parameter $a$ and the normalized mass variable $u = M_{\alpha}/r_e$. From the figure it is clear that considerable differences between Kerr and Kerr-MOG BHs will start being noticed when the ratio $u$ is at least of order $10^{-1}-10^{-2}$ since the grater the $u$, the bigger the difference between the predictions of both theories, confirming the fact that general relativistic effects and tests of the Kerr BH hypothesis will be more detectable in the vicinity of the BH horizon, where gravity is strong.

For completeness we have also plotted the difference of red- and blueshifts $\Delta z$ of photons emitted by corotating  and counterrotating particles (stars) around the Kerr-MOG black hole and the red- and blueshifts of photons corresponding to the Kerr BH in Fig. \ref{fig7_DShifts}. Again, from the figure we can infer that these differences point to the need of having $u$ at least of order $10^{-1}-10^{-2}$ in order to distinguish between a Kerr and a Kerr-MOG BH configuration. In this regard, the current available data for the S02 star orbiting around SgrA* yield a mass-to-radius ratio of order $u\sim 10^{-4}$, indicating that we need to improve our resolution to track orbits with at least $2$ orders smaller.

In this regard, a natural question is if a near to the event horizon orbit of a star with a sufficiently high velocity is still outside the distance from the BH beyond which stars are tidally disrupted. Since relativistic effects of supermassive BHs (related to its mass and spin) that tidally disrupt most of the main-sequence stars start being relevant when $M>10^7 M_{\odot}$ \cite{Kesden}, for SgrA* the Newtonian approach can be used in a good approximation. Thus, the Newtonian tidal disruption radius of a supermassive BH reads \cite{Hills-1975},\cite{Komossa-2015}
\begin{eqnarray}
\label{r_td}
r_{td} = 7 \times 10^{7} \left(\frac{M_{\textrm{BH}}}{10^6 M_{\odot}}\right)^{\frac{1}{3}} \left(\frac{M_{\star}}{M_{\odot}}\right)^{-\frac{1}{3}} 
\left(\frac{r_{\star}}{r_{\odot}}\right) \textrm{km},
\end{eqnarray}
where $M_{\star}$ and $r_{\star}$ denote the mass and radius of the star, respectively, and $M_{\textrm{BH}}$ is the mass of a supermassive black hole. By considering this relation, we can roughly estimate the Roche (tidal) radius and theruefore its normalized mass $u_{td}$ for different kinds of stars and see its effect on the measurement of their red- and blueshifts. For instance, for a neutron star we obtain $u_{td}\sim 5\times 10^3$; for a white dwarf we get $u_{td}\sim 5$, whereas for a star with a solar mass we arrive at $u_{td} \sim 5\times10^{-2}$. The first two cases correspond to orbits much closer to the event horizon of a supermassive BH compared to the S0 set of stars which revolve SgrA*; the case of a star like our Sun seems to be critical in the sense that it matches the region in which we need to track the star orbits with sufficiently high star velocities (or red-/blueshifts) to observe the desired effect. Notwithstanding, this Newtonian estimation is quite rough, a relativistic approach to this issue is much more complicated, there is not much literature available \cite{Kesden},\cite{Wiggins}, and it definitely deserves more attention.

On the other hand, in order to detect the aforementioned redshift differences for the S0 orbits around SgrA* that we have at hand now, we need a very precise spectrometer. However, nowadays the precision required by a spectrometer to measure these red- and blueshifts of photons, and therefore to constrain mass and spin-related quantities is not currently affordable with the facilities available at the moment. Hopefully, future instruments like MICADO at the ELT \cite{MICADO} will allow such measurements and therefore will allow us to perform tests of the Kerr BH hypothesis in the near future. One more project that has recently started is the STRONGGRAVITY EU FP7-SPACE \cite{STGRAVEU}; this facility will also aim to measure both spin and mass parameters of a BH through measurements of x-ray radiation, a region where the precision is considerably better in comparison with the infrared and radio regions.

We should mention as well that it is important to keep in mind the systematic uncertainties of the proposed method for determining the mass and spin parameters of the Kerr-MOG BH; this issue is mostly related to the emission and propagation
of photons in the astrophysical space, the orbital properties of the stars revolving the BH, and the limit in which stars can be considered geodesic particles, like in a binary system, among others (please see the discussion in Sec. VII).


\section{The mass and spin parameters of the Kerr-MOG black hole in terms of red- and blueshift of photons}

The mass $M_\alpha$ and spin $a$ parameters of the Kerr-MOG BH can be obtained from red- and blueshifts of the photons emitted by the source with the help of Eqs. (\ref{ZrKstatic}) and (\ref{ZbKstatic}) as it was pointed out in \cite{HN_2015}.

We first deduce an expression for the spin parameter in terms of the mass and the parameters of the Kerr-MOG BH metric:
\begin{eqnarray}
\label{a2}
a^2= \frac{\lambda r^{4}_{e}\left(r^{2}_{e}-2 M_{\alpha}r_{e}+M_{\beta}\right)}{\left(2 M_{\alpha} r_{e}-M_{\beta}\right)^{2}\eta-\lambda r^{4}_{e}},
\end{eqnarray}
where we have introduced the following quantities $\lambda=(z_{r}+z_{b})^{2}$ and $\eta=(z_{r}-z_{b})^{2}$.
The equation for the mass $M_\alpha$ calculated below from Eqs. (\ref{ZrKstatic})-(\ref{a2}) is of sixteenth order and cannot be solved analytically:
\begin{eqnarray}
&&\Big\{\left(r^{2}_{e}+2 M_{\beta}-3 M_{\alpha}r_{e}\right)\left(r^{2}_{e}-2 M_{\alpha} r +M_{\beta}\right)\times\nonumber\\
&&\left[\left(2 M_{\alpha}r_{e}\!-\!M_{\beta}\right)^{2}\eta-\lambda r^{4}_{e}\right]\!-\!4 r^{3}_{e}\varpi_e \left(2M_{\alpha}r_{e}\!-\!M_{\beta}\right)^{2} \Big\}^2\!=\nonumber\\
&&4\lambda r^{5}_{e}\varpi_e \left(r^{2}_{e}-2 M_{\alpha}r+M_{\beta}\right)^{3}\left[\left(2M_{\alpha}r_{e}\!-\!M_{\beta}\right)^{2}\eta-\lambda r^{4}_{e}\right]\!.
\nonumber\\
\label{M16}
\end{eqnarray}
However, even this form of the equation is useful for extracting the value of the mass parameter $M_\alpha$ in terms of the red- and blueshifts ($z_r,z_b$) and
the deformation parameter $\alpha$ by making use of a Bayesian estimation. Hence, to calculate the mass of a Kerr-MOG BH from astrophysical data one must use a statistical approach. For the case when $\alpha \rightarrow 0$, Eq. (\ref{M16}) becomes
\begin{eqnarray}
 \left[16 r_{e}M^3 \right.\!\!\!&-&\!\!\!\left. \left(4\eta M^2-\lambda r^{2}_{e}\right)(r_{e}-2M)(r_{e}-3M)\right]^2\nonumber\\
 &=& 4\lambda r^{2}_{e}M(r_{e}-2M)^3\left(4\eta M^2-\lambda r^{2}_{e}\right),
\label{M_Kerr}
\end{eqnarray}
which is the polynomial equation of the mass parameter $M$ of the Kerr BH as shown in \cite{HN_2015}.

Once we have statistically determined the value of the Kerr-MOG BH mass $M_\alpha$, then we should proceed to compute the spin parameter $a$ with the aid of the formula (\ref{a2}); both parameters will depend on alpha and they can give an estimation of departures from the predictions of the Kerr BH of the GTR.


\section{Conclusion and Discussion}

We studied the Kerr-MOG BH, which has an additional deformation parameter $\alpha$ with respect to the Kerr BH. This extra parameter $\alpha$, other than mass and spin parameters, directly influences the geometry of the Kerr-MOG BH as seen in Figs. \ref{fig1_gtt}, \ref{fig2_delta}, and \ref{fig5_ergo}. From Fig. \ref{fig5_ergo}, we concluded that with the increase in the deformation parameter $\alpha$ the ergoregion increases for given values of the spin parameter $a$, making the Kerr-MOG BH very interesting from the physical point of view.

We then showed in Fig. \ref{fig4_BH_NBH} the separation of the region in which the BH exists from the region where there is no BH. Hence, with this figure, we showed how the deformation parameter $\alpha$ restricts the value of the spin parameter $a$ of the Kerr-MOG BH. Thus, we concluded from Eq. (\ref{alphabounds1}), Fig. \ref{fig4_BH_NBH} and, Table \ref{table1} that the upper bound on parameter $\alpha$ decreases as the parameter $a$ increases.

We further studied the geodesic motion of both massive and massless particles in the gravitational field of the Kerr-MOG BH which was followed by the computation of the red- and blueshifts that emitted photons by massive geodesic particles experience when moving in the background of the Kerr-MOG spacetime.

We then computed the physical bounds on the red- and blueshifts of photons emitted by stars orbiting the Kerr-MOG BH by considering the static ($a=0$) and extremal ($|a|=M_{\alpha}/\sqrt{1+\alpha}$) limits. By comparing them to the Kerr BH metric predictions of GTR, we identified a novel simple test of the Kerr hypothesis in the strong gravitational regime. However, from the presented plots it is clear that considerable differences between Kerr and Kerr-MOG BHs will be noticeable when the mass-to-radius ratio $u$ is of order $10^{-1}-10^{-2}$; current available data for the S02 star orbiting around SgrA* yield a ratio of order $u\sim 10^{-4}$, indicating that we need to improve our resolution in at least $2$ orders and detect stars revolving closer to the BH. Another possibility that allows one to detect these differences is a very precise spectrometer. Notwithstanding, the required spectrometer precision to measure these red- and blueshifts of photons, and therefore to constrain BH mass and spin-related quantities is not currently available. Hopefully, future instruments like MICADO at the ELT and STRONGGRAVITY EU FP7-SPACE will allow such measurements and therefore will allow us to perform tests of the Kerr BH hypothesis in the near future.

Finally, we obtained a formula for the spin parameter $a$ in Eq. (\ref{a2}) in terms of the red- and blueshifts of photons emitted by the geodesic massive particles, the deformation parameter $\alpha$, and the radius of the massive objects orbiting around the Kerr-MOG BH emitting light. On the other hand, the mass parameter $M_{\alpha}$ was obtained as the root of the polynomial Eq. (\ref{M16}). The polynomial Eq. (\ref{M16}) in $M_{\alpha}$ is of sixteenth order, which cannot be solved analytically, but can be addressed statistically, which means that if we have an astronomical data set of red- and blueshifts of photons, the deformation parameter $\alpha$, and the orbital radii $r_e$ of different stars, we can compute the most possible value of the mass $M_{\alpha}$
 of the Kerr-MOG BH with a Bayesian fitting, for instance. Once the value for $M_{\alpha}$ is fixed, one can return to Eq. (\ref{a2}) in order to obtain a value for the spin parameter $a$.

We would like to point out as well that this analysis can be applied to astrophysical phenomena like active galactic nuclei and accretion disks, making it very important from the astrophysical point of view. Moreover, this method can also be implemented in several 4D BH solutions that have been obtained within the framework of higher-dimensional theories, like the stringy black hole considered in \cite{Uniyaletal, GGK}, for instance. Other stringy BH configurations that include a different amount of scalar parameters can be found in \cite{gk,hk3}. Moreover, our algorithm can be adapted to estimate the mass and spin of astrophysical BHs that have an accretion and disk around them. The recent research project of the European Space Agency named STRONGGRAVITY EU FP7-SPACE has come up with an objective to develop analytical tools to study the properties of black holes (like its mass \cite{mass} and spin \cite{spin}). Within this project the mass of the black hole can be measured from the accretion disk around it because the outer parts of the disk are cooler in comparison to its inner parts where the temperature can reach millions of degrees and the radiation coming from this part will fall in x-ray region; whereas the radiation coming from the outer parts of the disk fall in the UV and visible region. This radiation which falls in the visible region can be both red- and blueshifted depending upon the position of the accretion disk in a binary. If it is coming towards the observer the photons which are ejected from the disk are blueshifted and the photons are redshifted if the disk is receding from the observer. This is an interesting issue that we are currently addressing.

Here we should finally recall that, in general, the proposed method to compute the spin $a$ and mass $M$ parameters of a rotating BH in terms of observations relies on the use of a minimal set of assumptions; namely, the bodies that move around the BH are massive test particles that follow {\it stable geodesic orbits} and the photons they emit, propagate towards us along {\it null geodesics}. The so-far reported formulas for the red- and blueshifts correspond to particular circular equatorial orbits. The fact that stars follow stable geodesic orbits is a quite good approximation when considering stars revolving around a supermassive BH since the distance which separates them is huge, even when considering the central BH of our Galaxy; however, quite often we find highly eccentric orbits that lie out of the equatorial plane, like the stars orbiting around SgrA*, these orbits require a refinement of the method that considers these orbital properties and is a topic in which we are currently working.

On the other hand, photons can (and most probably do) be absorbed and reemitted during their propagation in the astrophysical medium from the moment of emission till detection. We are also studying the possibility of incorporating this effect in our approach; another possible exit from this trouble is to subtract this effect from the observations. This problem gets worse when considering accretion disks since their complicated magnetohydrodynamic processes may strongly influence the photon dynamics and therefore may impact the relation between their red-/blueshifts and the black hole parameters. The application of this method to BHs accreting gas is an interesting open problem that definitely deserves more attention. Another interesting project consists of applying a suitable version of this method to binary systems [either BHs or neutron stars (NSs) or a BH with a NS]. However, here there are two major caveats that should be taken into account; namely, the masses of the two orbiting objects can be comparable to each other, and hence there will be no test particle approximation, and the orbits of the bodies can be nongeodesic due to their size. This issue is also under current research.

\section*{APPENDIX: DETAILED CALCULATION OF
CONSERVED QUANTITIES}

Here we shall derive with some detail the relations (\ref{E1}) and (\ref{L1}) by starting from the conditions (\ref{Vmin}):
\begin{eqnarray}
V_r & = &T^2 - \left(r^2 - 2M_{\alpha}r + a^2 + M_{\beta}\right)\times\nonumber\\
& &\left[r^2+\left(L-aE\right)^2\right] = 0,
\label{Vr}
\end{eqnarray}
\begin{eqnarray}
&&V'_r = 2ET - \left(1 - \frac{M_{\alpha}}{r}\right)\left[r^2+\left(L-aE\right)^2\right] - \nonumber\\
&&\left(r^2 - 2M_{\alpha}r + a^2 + M_{\beta}\right) = 0.
\label{V1r}
\end{eqnarray}
In terms of the reciprocal radius $u=1/r$ these equations reduce to
\begin{eqnarray}
&&\left[E-a\left(L-aE\right)u^2\right]^2 - \left[1+\left(L-aE\right)^2u^2\right]\times\nonumber\\
&&\left(1 - 2M_{\alpha}u + a^2u^2 + M_{\beta}u^2\right) = 0,
\label{Vu}
\end{eqnarray}
\begin{eqnarray}
&&2E\Big[E\!-\!a\left(L\!-\!aE\right)\!u^2\Big]\!-\!\left(1\!-\!M_{\alpha}u\right)\left[1\!+\!\left(L\!-\!aE\right)^2\!u^2\right]\!-\nonumber\\
&&\left(1\!-\!2M_{\alpha}u + a^2u^2 + M_{\beta}u^2\right) = 0.
\label{V1u}
\end{eqnarray}
By subtracting one expression from the other we can obtain the following expression
\begin{eqnarray}
E^2 = \left(1-M_{\alpha}u\right)+x^2\left(M_{\alpha}-M_{\beta}u\right)u^3,
\label{E2}
\end{eqnarray}
where we have introduced $x = L-aE$. By inserting this expression into (\ref{V1u}) we obtain
\begin{eqnarray}
&&2aEux = M_{\alpha}-a^2u-M_{\beta}u+\nonumber\\
&&x^2\left(-1+3M_{\alpha}u-2M_{\beta}u^2\right)u,
\label{Euax}
\end{eqnarray}
By equating Eqs. (\ref{E2}) and (\ref{Euax}) we obtain a quadratic equation for $x^2u$:
\begin{eqnarray}
&&\left[\left(1-3M_{\alpha}u +2M_{\beta}u^2\right)^2\!-\!4a^2(M_{\alpha}-M_{\beta}u)u^3\right]x^4u^2 - \nonumber\\
&&2\left[\left(1-3M_{\alpha}u +2M_{\beta}u^2\right)
\left(M_{\alpha}-a^2u-M_{\beta}u\right) + \right.\nonumber\\
&&\left.2a^2\left(1-M_{\alpha}u\right)u\right]x^2u+\left(M_{\alpha}-a^2u-M_{\beta}u\right)^2=0,
\label{x2u}
\end{eqnarray}
with the following discriminant
\begin{eqnarray}
&&\left(b/2\right)^2-ac=4a^2(M_{\alpha}-M_{\beta}u)u\times\nonumber\\
&&\left[1-(2M_{\alpha}-M_{\beta}u)u + a^2u^2\right]^2,
\label{discriminant}
\end{eqnarray}
which, upon multiplication by $u$ and some algebraic manipulations, allows us to write the solution as
\begin{equation}
x^2u^2 = \frac{Q_{\pm}\Delta_u-Q_{+}Q_{-}}{Q_{+}Q_{-}}=\frac{\Delta_u-Q_{\mp}}{Q_{\mp}},
\label{x2u2}
\end{equation}
where we have introduced the following quantities
\begin{eqnarray}
&&Q_{\pm} = 1-3M_{\alpha}u +2M_{\beta}u^2\pm \nonumber\\
&&2au\sqrt{(M_{\alpha}-M_{\beta}u)u},
\label{Q}
\end{eqnarray}
\begin{equation}
\Delta_u=1-(2M_{\alpha}-M_{\beta}u)u + a^2u^2.
\label{D}
\end{equation}
Remarkably, the numerator in (\ref{x2u2}) can be written as a squared quantity
\begin{equation}
\Delta_u-Q_{\mp}=\left(au\pm \sqrt{(M_{\alpha}-M_{\beta}u)u}\right)^2,
\label{DQ}
\end{equation}
enabling us to get a simple expression for the $x$ variable:
\begin{equation}
x=-\frac{a\sqrt{u}\pm\sqrt{M_{\alpha}-M_{\beta}u}}{\sqrt{uQ_{\mp}}},
\label{x}
\end{equation}
where one should notice that only the minus sign satisfies the condition (\ref{V1r}). We further insert this solution into Eq. (\ref{E2}) and note that the rhs can be written as a squared quantity as well, leading to the following expression for $E$,
\begin{eqnarray}
E\!=\!\frac{1}{\sqrt{Q_\mp}}\left[1\!-\!(2M_{\alpha}\!-\!M_{\beta}u)u\!\mp\!
au\sqrt{(M_{\alpha}\!-\!M_{\beta}u)u}\right]
\label{Ealpha}
\end{eqnarray}
and by finally making use of the $x$ definition we obtain for $L$:
\begin{eqnarray}
&&L = \frac{1}{\sqrt{uQ_\mp}}\Big[\mp\sqrt{M_{\alpha}-M_{\beta}u}\,(1+a^2u^2)-\nonumber\\
&&(2M_{\alpha}\!-\!M_{\beta}u)au^{3/2}\Big].
\label{Lalpha}
\end{eqnarray}
It is straightforward to check that expressions (\ref{Ealpha})and (\ref{Lalpha}) coincide with (\ref{E1})\textendash(\ref{pi}).


\section*{Acknowledgements}
P.S. is grateful to Professor S. Kalyan Rama for warm hospitality at the Institute of Mathematical Sciences, Chennai, India where part of this work was performed. He would also like to thank Professor Sushant G. Ghosh and Professor Sanjay Siwach for continuous encouragement. In particular the authors acknowledge fruitful correspondence with Professor John W. Moffat that was helpful in understanding the Kerr-MOG metric. They are grateful  to Ra\'ul Lizardo-Castro and Bryan Larios for interesting discussions. The work of A.H.-A. was completed at the Aspen Center for Physics, which is supported by National Science Foundation Grant No. PHY-1607611 and a Simons Foundation grant as well. He expresses his gratitude to the ACP for providing an inspiring and encouraging atmosphere for conducting part of this research. A.H.-A. and U.N. acknowledge support from SNI and CONACYT, PRODEP, VIEP-BUAP and CIC-UMSNH. U.N. thanks the CONACYT thematic network Project No. 280908 {\it "Agujeros Negros y Ondas Gravitatorias"} for financial support.



\begin{thebibliography}{99}
\bibitem{footnote1}
In previous versions of this metric, Newton's gravitational constant was believed to be enhanced through the relation $G=G_{N}(1+\alpha)$ appearing in the metric function $\Delta$. However, by computing the ADM mass of this metric we find that the factor $(1+\alpha)$ affects the mass, i.e. $M_{ADM}=M(1+\alpha)$ [the rhs is written in the language in which the metric (\ref{metric}) is presented in \cite{Moffat_2015}; see below], leading the Newton's constant unmodified, i.e., $G=G_N$.
\bibitem{footnote2}
The parameter $M$ is only an integration constant of the Einstein equations and it cannot be interpreted as the black hole mass since it is not the value of the Hamiltonian for the metric (\ref{metric}).
\end{thebibliography}
\end{document}